\documentclass[a4paper,twocolumn]{article} 

\usepackage[top=1.5cm, bottom=1.5cm, left=1.5cm, right=1.5cm, bindingoffset = 0pt, footskip = 0.75cm, marginparwidth = 0pt, marginparsep=0pt]{geometry}

\usepackage{helvet}

\usepackage{comment}

\usepackage{upgreek}

\usepackage{graphicx}
\usepackage{float}
\usepackage{caption}
\usepackage{amsmath}
\usepackage{amsfonts}
\usepackage{amssymb}

\usepackage[switch]{lineno}

\usepackage{etoolbox} 
\makeatletter 
\newcommand*\linenomathpatch{\@ifstar{\linenomathpatch@AMS}{\linenomathpatch@}}
\newcommand*\linenomathpatch@[1]{
  \expandafter\pretocmd\csname #1\endcsname {\linenomathWithnumbers}{}{}
  \expandafter\pretocmd\csname #1*\endcsname{\linenomathWithnumbers}{}{}
  \expandafter\apptocmd\csname end#1\endcsname {\endlinenomath}{}{}
  \expandafter\apptocmd\csname end#1*\endcsname{\endlinenomath}{}{}
}
\newcommand*\linenomathpatch@AMS[1]{
  \expandafter\pretocmd\csname #1\endcsname {\linenomathWithnumbersAMS}{}{}
  \expandafter\pretocmd\csname #1*\endcsname{\linenomathWithnumbersAMS}{}{}
  \expandafter\apptocmd\csname end#1\endcsname {\endlinenomath}{}{}
  \expandafter\apptocmd\csname end#1*\endcsname{\endlinenomath}{}{}
}
\let\linenomathWithnumbersAMS\linenomathWithnumbers
\patchcmd\linenomathWithnumbersAMS{\advance\postdisplaypenalty\linenopenalty}{}{}{}
\makeatother 

\linenomathpatch{equation}
\linenomathpatch*{gather}
\linenomathpatch*{multline}
\linenomathpatch*{align}
\linenomathpatch*{alignat}
\linenomathpatch*{flalign}

\usepackage{colortbl}

\usepackage[style=nature]{biblatex}
\newcommand{\eq}[1]{(\ref{#1})}
\newcommand{\cov}{\mathrm{cov}}
\newcommand{\var}{\mathrm{var}}

\DeclareMathOperator*{\argmin}{arg min}

\title{Non-Gaussian noise spectroscopy with a superconducting qubit sensor}

\usepackage{authblk}
\author[1,2]{Youngkyu Sung}
\author[3]{F\'elix Beaudoin}
\author[3]{Leigh M. Norris}
\author[1]{Fei Yan}
\author[4]{David K. Kim}
\author[1,2]{Jack Y. Qiu}
\author[1]{Uwe von L\"upke}
\author[4]{Jonilyn L. Yoder}
\author[1,2]{Terry P. Orlando}
\author[1]{Simon Gustavsson}
\author[3]{Lorenza Viola \thanks{lorenza.viola@dartmouth.edu}}
\author[1,2,4,5]{\mbox{William D. Oliver} 
\thanks{ william.oliver@mit.edu}}

\affil[1]{Research Laboratory of Electronics, Massachusetts Institute of Technology, Cambridge, MA 02139, USA}
\affil[2]{Department of Electrical Engineering and Computer Science, Massachusetts Institute of Technology, Cambridge, MA 02139, USA}
\affil[3]{Department of Physics and Astronomy, Dartmouth College, Hanover, NH 03755, USA}
\affil[4]{MIT Lincoln Laboratory, 244 Wood Street, Lexington, MA 02421, USA}
\affil[5]{Department of Physics, Massachusetts Institute of Technology, Cambridge, MA 02139, USA}

\usepackage{times}
\date{\today}

\begin{document}


\maketitle

\textsf{\bfseries{ 
Accurate characterization of the noise influencing a quantum system of interest has far-reaching 
implications across quantum science, ranging from microscopic modeling of decoherence dynamics to noise-optimized quantum control. While the assumption that noise obeys Gaussian statistics is commonly employed, noise is generically non-Gaussian in nature. In particular, the Gaussian approximation breaks down whenever a qubit is strongly coupled to discrete noise sources or has a non-linear response to the environmental degrees of freedom. Thus, in order to both scrutinize the applicability of the Gaussian assumption and capture distinctive non-Gaussian signatures, a tool for characterizing non-Gaussian noise is essential. Here, we experimentally validate a quantum control protocol which, in addition to the  spectrum, reconstructs the leading higher-order spectrum of engineered non-Gaussian dephasing noise using a superconducting qubit as a sensor. This first experimental demonstration of non-Gaussian noise spectroscopy represents a major step toward demonstrating a complete spectral estimation toolbox for quantum devices.
}}

\section*{Introduction}

For any dynamical system that evolves in the presence of unwanted disturbances, precise knowledge of the noise spectral features is fundamental for quantitative understanding and prediction of the dynamics under realistic conditions. As a result, spectral estimation techniques have a long tradition and play a central role in classical statistical signal processing \cite{percival1993spectral}. For quantum systems, the importance of precisely characterizing noise effects is further heightened by the challenge of harnessing the practical potential that quantum science and technology applications promise. Such detailed knowledge is key to develop noise-optimized strategies for enhancing quantum coherence and boosting control fidelity in near-term intermediate-scale quantum information processors \cite{NISQ}, as well as for overcoming noise effects in quantum metrology \cite{Dur2016,Felix2018}. Ultimately, probing the extent and decay of noise correlations will prove crucial in determining the viability of large-scale fault-tolerant quantum computation \cite{Preskill2013}. 

Thanks to their exquisite sensitivity to the surrounding environment, qubits driven by external control fields are naturally suited as ``spectrometers,'' or sensors, of their own noise \cite{Schoelkopf2002,Faoro2004}. Quantum noise spectroscopy (QNS) leverages the fact that open-loop control modulation is akin to shaping the filter function that determines the sensor's response in frequency space \cite{FF2008,Biercuk2009,Yuge2011,Young2012,Paz-Silva2014} and, in it simplest form, aims to characterize the spectral properties of environmental noise as sensed by a single qubit sensor. By now, QNS protocols employing both pulsed and continuous control modalities have been explored, and experimental implementations have been reported across a wide variety of qubit platforms -- including NMR~\cite{Alvarez2011}, superconducting quantum circuits~\cite{Bylander2011a,Yan2013,Yoshihara2014,Quintana2017}, semiconductor quantum dots~\cite{Dial2013,Muhonen2014,Morello2018,Tarucha2018}, diamond nitrogen vacancy centers~\cite{Meriles2010,Romach2015}, and trapped ions \cite{Frey2017}. Notably, knowledge of the underlying noise spectrum has already enabled unprecedented coherence times to be achieved via tailored error suppression \cite{Kim2017}. 
 
While the above advances clearly point to the growing significance of spectral estimation in the quantum setting, they all rely on the assumption that the target noise process is Gaussian -- that is, one- and two-point correlation functions suffice to fully specify the noise statistical properties. However, the Gaussian assumption need not be justified {\em a priori} and it should rather be validated (or falsified) by the QNS protocol itself. A number of realistic scenarios motivate the consideration of non-Gaussian noise regimes. Statistical processes that are responsible for electronic current fluctuations in mesoscopic devices or the 1/$f$ noise ubiquitously encountered in solid-state quantum devices are not Gaussian in general~\cite{Paladino2014a}. In superconducting circuits, previous studies have shown that a few two-level defects within Josephson tunnel junctions can interact strongly with the qubit \cite{Simmonds2004,Oliver2013,PalmerPRB2013,Lisenfeld2015,Lisenfeld2016}, the resulting decoherence dynamics showing marked deviations from Gaussian behavior under both free evolution and dynamical decoupling protocols \cite{Faoro2004,FalciPRA2004,GalperinPRB2007}. More generally, non-Gaussian noise statistics may be expected to arise whenever a qubit is operated outside a linear-response regime, either due to strong coupling to a discrete environment \cite{Kotler2013} or to a non-linear energy dispersion relationship. The latter feature, which has long been appreciated to influence dephasing behavior at optimal points  \cite{Makhlin2004}, is common to all state-of-the-art superconducting qubit archetypes \cite{Barends2014,Yan2016,Hutchings2017,Lin2018}. Thus, statistical correlations higher than second order and their corresponding multi-dimensional Fourier transforms must be taken into account for complete characterization. From a signal processing standpoint, this translates into the task of higher-order spectral estimation \cite{Nikias1993}.

In this work, we experimentally demonstrate non-Gaussian QNS by building on the estimation procedure proposed by Norris {\em et al.}~\cite{Norris2016}. While we employ a flux-tunable superconducting qubit as a sensor, our methodology is portable to other physical testbeds in which classical dephasing noise is the dominant decoherence mechanism. We show how non-Gaussianity distinctively modifies the phase evolution of the sensor's coherence, resulting in an observable signature to which the spectrum (or power spectral density, PSD) is completely insensitive and which is instead encoded in the leading higher-order spectrum, the bispectrum. Unlike the original proposal~\cite{Norris2016}, the QNS protocol we introduce here makes use of a statistically-motivated maximum likelihood approach. This renders the estimation less susceptible to numerical instability, while allowing measurement errors to be incorporated and both the PSD and the bispectrum to be inferred using a single measurement setup. In order to obtain a clean benchmark for our spectral estimation procedure, we engineer a non-Gaussian noise model by injecting Gaussian flux at the sensor's degeneracy point, resulting in non-Gaussian frequency noise. The noise implementation is validated by verifying the observed power dependence of the leading cumulants against the expected one. Both the reconstructed PSD and the bispectrum are found to be in quantitative agreement with theoretical predictions within error bars. 

\section*{Results}

\subsection*{Non-Gaussian dephasing noise}
Before introducing our experimental test bed, we present the general setting to which our analysis is relevant: a qubit sensor evolving under the combined action of non-Gaussian classical dephasing noise and suitably designed sequences of control pulses.  By working in an interaction frame with respect to the internal qubit Hamiltonian and the applied control, and letting $\hbar=1$, the controlled open-system Hamiltonian may be written as $H (t)=y_p(t) B(t) \sigma_z/2,$ where $B(t)$ is a stochastic process describing dephasing noise relative to the qubit's eigenbasis defined by the Pauli operator $\sigma_z$. The control switching function $y_p(t)$ accounts for a sequence $p$ of instantaneous $\pi$ rotations about the $x$ or $y$ axis, starting from initial value $y_p(0)=+1$ and toggling between $\pm1$ with every application of a pulse. Under such a pure-dephasing Hamiltonian, the qubit coherence is quantified by the time-dependent expectation value $\langle\sigma_+(t)\rangle \equiv \mathrm e^{-\chi(t)+i\phi(t)}\langle \sigma_+(0)\rangle$, where the influence of the noise is captured by the decay and phase parameters $\chi(t)$ and $\phi(t)$. These parameters may be formally expanded in terms of noise cumulants, $C^{(k)}(t_1,\ldots,t_k)$, $k \, \in\, \{1,2,\ldots,\infty\}$, with $\chi(t)$ taking contribution only from even cumulants and $\phi(t)$ only from odd cumulants \cite{Norris2016}. Physically, the $k$-th order cumulant is determined by the multi-time correlation functions $\mathbb{E}[B(t_1),\ldots,B(t_j)]$, with $j\leq k$, where $\mathbb{E}[\cdot]$ denotes the ensemble average over noise realizations. 

Since the statistical properties of Gaussian noise are entirely determined by one- and two-point correlation functions, cumulants of order $k\geq 3$ vanish identically. By contrast, for non-Gaussian noise, all cumulants can be non-zero in principle. Assuming that noise is stationary, so that the mean of the process, $\mathbb{E}[B(t)]=C^{(1)}(0)\equiv \mu_B$ is constant, the phase parameter may be written as $\phi(t)=\mu_B F_p(0,t)+\varphi(t)$, with the Fourier transform $F_p(\omega,t)\equiv\int_0^t d\mathrm{s}\,\mathrm e^{-i\omega s}y_p(s)$ being the fundamental filter function (FF) associated to the control \cite{Paz-Silva2014}. This expression separates the phase due to the noise mean, which arises for both Gaussian and non-Gaussian noise, from a genuinely \textit{non-Gaussian phase} $\varphi(t)$, which captures the contribution of all odd noise cumulants with $k\geq3$. For sufficiently small time or noise strength, we can neglect terms of order $k>3$ in the cumulant expansion, leading to
\begin{align}
\chi(t) &\approx \frac1{2\pi}\int_{\mathbb R} \mathrm{d}\omega |F_p(\omega,t)|^2S(\omega) ,	
\label{eqnchi}\\
\varphi(t) &\approx -\frac1{3!(2\pi)^2}\int_{\mathbb R^2} \mathrm{d}\vec \omega \,G_p(\vec\omega,t)S_2(\vec\omega),	
\label{eqnphi}
\end{align}
where $\vec \omega\equiv(\omega_1,\omega_2)$ and the second and third noise cumulants enter the qubit dynamics through their Fourier transforms: the PSD or spectrum, $S(\omega)\equiv\int_{\mathbb{R}} \mathrm{d}\tau \,\mathrm e^{-i\omega \tau} C^{(2)}(0,\tau)$, and the second-order polyspectrum or bispectrum, $S_2(\vec\omega)\equiv\int_{\mathbb R^2}\mathrm{d} \vec{\tau}\,\mathrm e^{-i\vec{\omega}\cdot\vec{\tau}}C^{(3)}(0,\tau_1,\tau_2)$, with $\vec \tau\equiv(\tau_1,\tau_2)$. In the frequency domain, the influence of such spectra is ``filtered'' by a corresponding generalized FF -- in particular, $G_p(\vec\omega,t)\equiv F_p(-\omega_1,t)F_p(-\omega_2,t)F_p(\omega_1+\omega_2,t)$ \cite{Paz-Silva2014}. Since, to leading order, non-Gaussian features arise in our setting from $S_2(\vec{\omega})$, non-Gaussianity of a noise process will be detected and characterized through measurements of $\varphi(t)$.

\begin{figure}[t!]
\centering
\includegraphics[width=8.9cm]{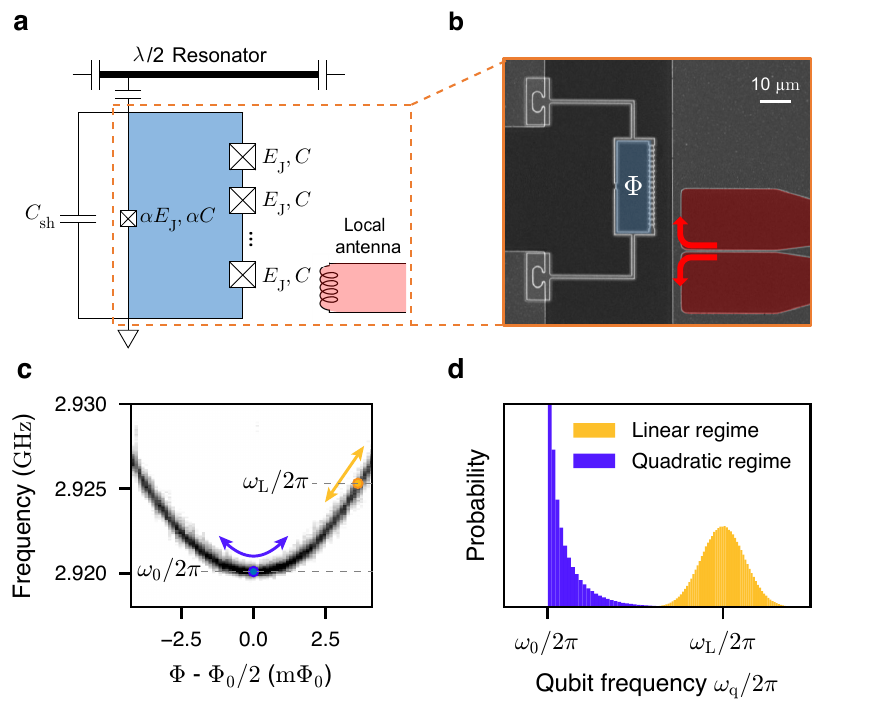}
\caption{\textbf{Experimental setup and non-Gaussian dephasing noise in a superconducting qubit}.
\textbf{a}, Schematic of the circuit QED system. An engineered flux qubit comprises a superconducting loop (blue) interrupted by one small-area and 8 large-area Josephson junctions (crosses) and is inductively coupled to a local antenna (red). The qubit junctions have internal capacitance, $C$ and $\alpha C$, and are externally shunted by capacitance $C_{\mathrm{sh}}$. See Supplementary Note~
1.
\textbf{b}, SEM image of the device. The flux threading the qubit loop $\Phi$ is modulated by applying a current through the local antenna.
\textbf{c}, Frequency spectroscopy of the qubit's $|0\rangle \rightarrow |1\rangle$ transition.
At (away from) the degeneracy point $\Phi=\Phi_0/2$, the qubit frequency $\omega_{\mathrm q}$ has a quadratic (linear) dependence on the external flux, as indicated by the indigo (yellow) arrow.
\textbf{d}, Probability distribution of the qubit frequency under Gaussian flux noise in the linear regime (yellow) vs. the quadratic regime (indigo). In the quadratic regime, the right-skewness of the distribution illustrates the non-Gaussianity of the resulting noise process.}
\label{fig:Fig1}
\end{figure}

\subsection*{Experimental setup and noise validation}

Our circuit QED system~\cite{Blais2004,Wallraff2004}  contains an engineered flux qubit~\cite{Yan2018}, which is designed to enable fast single-qubit gates with high fidelity at its flux degeneracy point ($F_g > 99.9\%$; see Supplementary Notes~
1 and 2). Single-qubit operations are performed using cosine-shaped microwave pulses, applying an optimal-control technique to suppress leakage to higher levels~\cite{Motzoi2009}. Inductive coupling to a local antenna is used to modulate the external flux $\Phi$ threading the qubit loop interrupted by Josephson junctions (Fig.~\ref{fig:Fig1}a and Fig.~\ref{fig:Fig1}b). Near the degeneracy (or optimal \cite{Makhlin2004}) point $\Phi = \Phi_0/2$, with $\Phi_0$ the flux quantum, the $|0\rangle \rightarrow |1\rangle$ transition frequency $\omega_{\mathrm q}$ has an approximately quadratic dependence on the external flux $\Phi$ (Fig.~\ref{fig:Fig1}c). Hence, a sufficiently slow time-dependent external flux $\Phi(t)$ enables adiabatic modulation of the qubit frequency, leading to 
\begin{align}
B(t)=\beta_\Phi\,[\Delta\Phi(t)]^2, \hspace{5mm}\Delta\Phi(t)\equiv \Phi(t)-\Phi_0/2,   
\label{eqnB}
\end{align}
where $\beta_\Phi$ is the quadratic coefficient in the dispersion relation between qubit frequency and flux. Crucially, any non-linear function of a Gaussian process leads to non-Gaussian noise. In particular, the quadratic function implemented in Eq.~\eq{eqnB} transduces 
zero-mean Gaussian flux noise 
into non-Gaussian qubit-frequency noise (Figs.~\ref{fig:Fig1}c and d). Assuming that the noise is entirely contributed by the applied $\Delta\Phi(t)$, and that $S_\Phi(\omega)$ denotes the corresponding PSD, the mean $\mu_B$, PSD $S(\omega)$, and bispectrum $S_2(\omega_1,\omega_2)$ of $B(t)$ are respectively given by 
\begin{align}
\mu_B&=\frac{\beta_\Phi}{2\pi}\int_{\mathbb R} \!\!\mathrm{d}\omega\, S_\Phi(\omega),	
\label{eqnmuB}\\
S(\omega)&=\frac{\beta_\Phi^2}{\pi}\int_{\mathbb R}\!\! d\mathrm{u}\, S_\Phi(u)S_\Phi(\omega-u),	
\label{eqnS}\\
S_2(\omega_1,\omega_2)&=\frac{4\beta_\Phi^3}{\pi}\int_{\mathbb R}\!\!\mathrm{d}u\,S_\Phi(u)S_\Phi(\omega_1+u)S_\Phi(\omega_2-u),	
\label{eqnS2}
\end{align}
In the experiment, we choose $S_\Phi(\omega)$ to be a zero-mean Lorentzian function, $S_{\Phi}(\omega) = (P_0/\pi \omega_c) /[1+(\omega / \omega_c)^2]$, where $\omega_c/2\pi$ $(=0.5~\mathrm{MHz})$ and $P_0$ denote the cutoff frequency and the power of the applied flux noise, respectively. As is apparent from Eqs.~\eq{eqnmuB}--\eq{eqnS2}, cumulants of order $k=1$, 2, and 3 are distinguished by their linear, quadratic, and cubic dependence on power, respectively.


\begin{figure}[t]
\centering
\includegraphics[width=8.9cm]{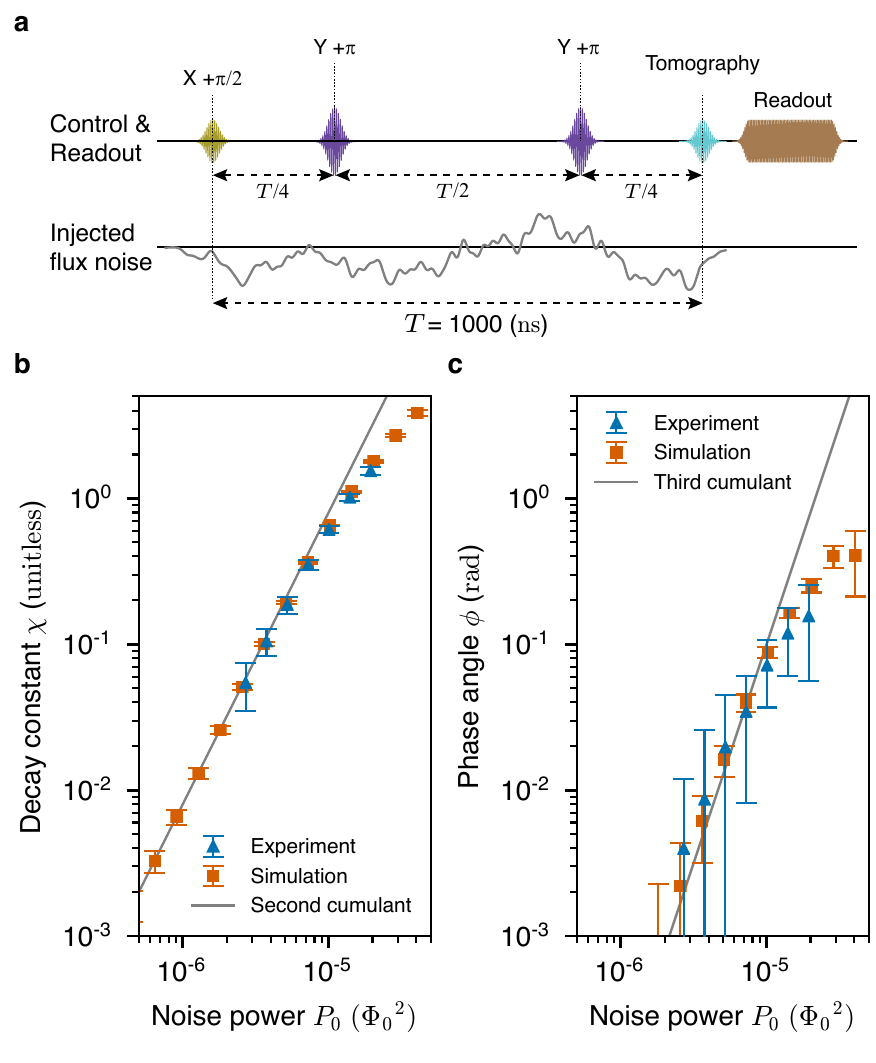}
\caption{\textbf{Power dependence of decay constant ($\chi$) and phase angle ($\phi$).}
\textbf{a}, Pulse scheme for measuring the power dependence of $\chi$ and $\phi$, consisting of a CPMG sequence of length $T=1~\mathrm{\mu s}$ with two $\pi$ pulses. Flux noise waveforms are temporally tailored to affect the qubit only while it evolves on the transverse plane.
\textbf{b}, Decay constant $\chi = -\log(\sqrt{\langle \sigma_x \rangle^2+\langle \sigma_y \rangle^2})$ and
\textbf{c}, phase angle $\phi=\tan^{-1}(- \langle \sigma_x \rangle / \langle \sigma_y \rangle)$ at time $t=T$, after application of a CPMG sequence as a function of the applied noise power $P_0$. A cubic power dependence of $\phi$, for sufficiently weak noise, corroborates non-Gaussianity of the engineered noise. Error bars represent 95\% confidence intervals.}
\label{fig:Fig2}
\end{figure}

We first validate the intended engineered non-Gaussian noise by demonstrating consistency of the measured power dependence of $\chi$ and $\phi$ with the above prediction. The qubit is initialized to the $+y$ axis by applying a $\pi/2$ pulse about $x$ (rotation $R_x(\pi/2)$), and Gaussian flux noise is injected while it evolves in the $xy$-plane of the Bloch sphere for time $T$. During this evolution, we apply a Carr-Purcell-Meiboom-Gill (CPMG) sequence consisting of two refocusing $\pi$ pulses about $y$ (Fig.~\ref{fig:Fig2}a). At the end of this sequence ($t=T$), the effect of the first cumulant of the noise cancels out ($F_p(0,T)=0$) and, as a result, the measured phase becomes solely determined by odd cumulants of order $k\geq3$: $\phi(T)=\varphi(T)$. To estimate both $\phi$ and $\chi$, we measure  $\langle\sigma_x\rangle$ and $\langle\sigma_y\rangle$ by applying appropriate tomography pulses at time $t=T$, before readout in the $\sigma_z$-basis.  

Figures~\ref{fig:Fig2}b and \ref{fig:Fig2}c show $\chi$ and $\phi$ as a function of injected flux noise power $P_0$ for both the experiment (blue triangles) and Monte Carlo simulations accounting for all cumulants of the applied noise (orange squares, see Supplementary  Note~
5). Substituting Eqs.~\eq{eqnS} and \eq{eqnS2} into Eqs.~\eq{eqnchi} and \eq{eqnphi}, we also plot the resulting ideal weak-power behavior (gray solid) considering only the leading-order cumulants of order two and three for $\chi$ and $\phi$, respectively. For sufficiently small $P_0$, these ideal values are in good agreement with data from both experiment and simulation, showing that $\chi$ and $\phi$ obey the quadratic and cubic power dependences that are expected for the square of a Gaussian flux-noise process under the CPMG sequence. In particular, the cubic dependence of $\phi$ at small $P_0$ corroborates the presence of a non-zero third-order cumulant, which would not exist for Gaussian noise. Deviations of the simulations and experimental data from the ideal behavior at large $P_0$ are attributable to the contribution of cumulants of order $k>3$. The quantitative agreement between theory, experiment, and simulation observed at low power demonstrates our capability to produce and sense engineered noise that dominates over native one over the relevant parameter regime and exhibits well controlled cumulants, a necessary first step in the experimental validation of non-Gaussian QNS.

\begin{figure}[h!]
\centering
\includegraphics[width=8.9cm]{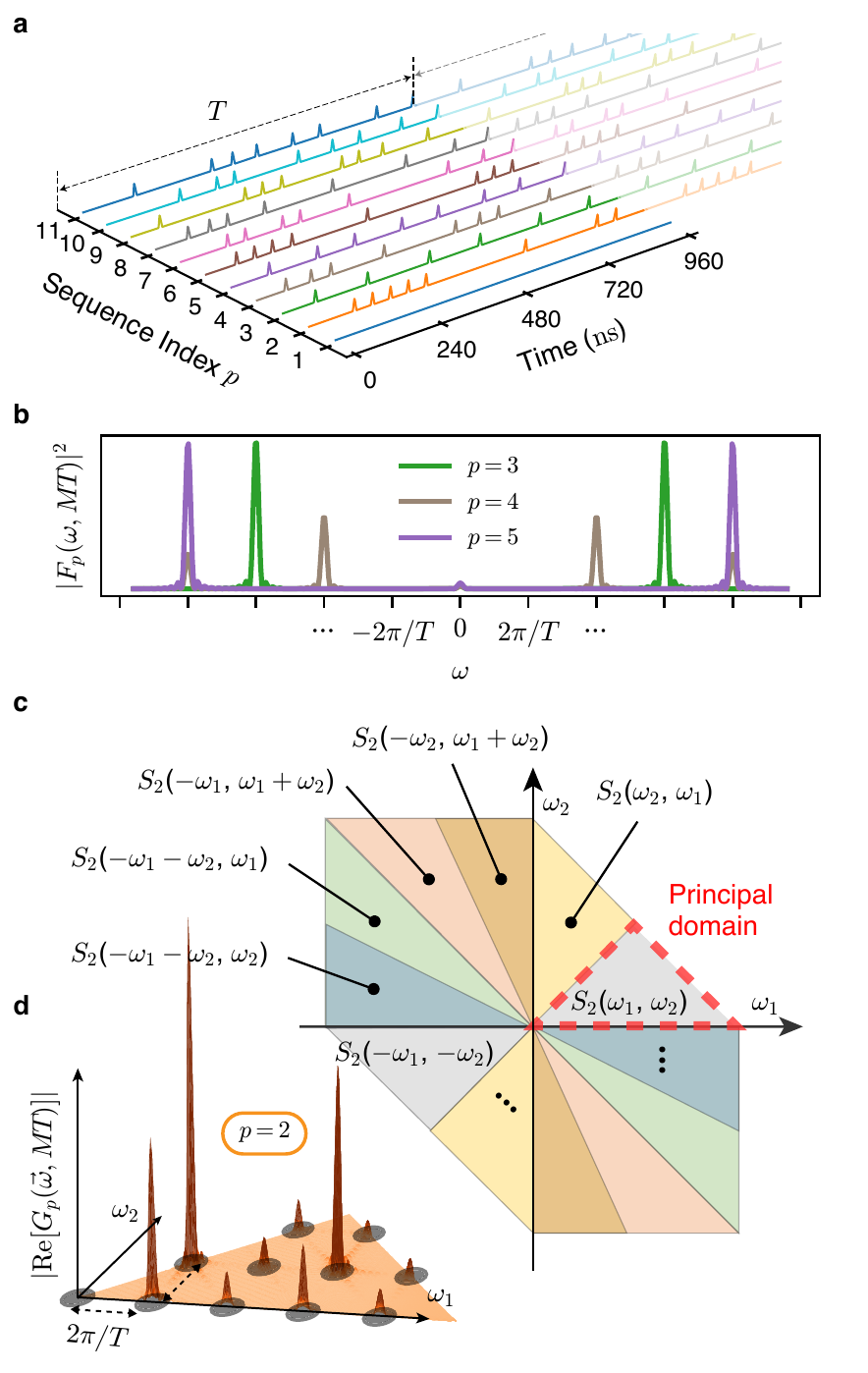}
\caption{\textbf{A protocol for non-Gaussian noise spectroscopy.}
\textbf{a}, Timing diagrams of control pulse sequences. The length of the base sequence is $T= 960$ $\mathrm{ns}$, $p=1$ corresponds to a single free-evolution period, whereas sequences $p=2, \dots, 11$ are repeated $M=10$ times. Only $\pi$-pulses are shown and all $\pi$-pulses are around the $y$-axis (see Supplementary Note~
4 for details).
\textbf{b}, $|F_p(\omega, MT)|^2$ for $p = 3, 4, 5$ as a function of angular frequency $\omega$.  
\textbf{c}, Symmetries of the bispectrum of a classical stationary noise process.
\textbf{d}, 2D grid representing the harmonic frequencies (black circles) in the principal domain ${\cal D}_2$ (orange area) in which the bispectrum is sampled. The amplitude of the relevant contribution of the FF in ${\cal D}_2$, $|{\rm Re}[G_p(\vec\omega, MT)]|$, for $p = 2$, (red surface plot) is shown on top of the grid. }
\label{fig:Fig3}
\end{figure}

\subsection*{Non-Gaussian noise spectroscopy}
Having established that $\chi$ and $\phi$ follow their expected behavior, we move on to fully characterizing the first three cumulants of our engineered noise source by measuring its mean, PSD, and bispectrum. Since the noise mean, $\mu_B$, manifests itself through a qubit-frequency shift, it can be measured from a simple parameter estimation scheme based on Ramsey interferometry. By contrast, we aim to perform a non-parametric estimation of both the PSD and bispectrum, that is, to reconstruct them at a set of discrete points in frequency space without assuming a prior functional form. Figure~\ref{fig:Fig3} illustrates our protocol for simultaneous estimation of the PSD and bispectrum, in which filter design -- the selection of pulse times in a control sequence so that the corresponding FF has a particular shape -- is instrumental. Building on Ref.~\cite{Alvarez2011}, applying $M\gg1$ repetitions of a ``base'' pulse sequence $p \in \{1,2, \cdots, P\}$, with duration $T$,  shapes the FF $|F_p(\omega,MT)|^2$ into a frequency comb with narrow teeth probing $S(\omega)$ at harmonics $k\omega_h$, with $k$ an integer and $\omega_h\equiv 2\pi/T$ (Figs.~\ref{fig:Fig3}a and b). This result generalizes to filters relevant to higher-order spectra \cite{Paz-Silva2014}: under sequence repetition, $G_p(\vec \omega,MT)$ becomes a two-dimensional (2D) ``hyper-comb'' with teeth probing $S_2(\vec \omega)$ at $\vec\omega\,\in\,\{\vec k\omega_h\}$, where $\vec k\equiv(k_1,k_2)$ with $k_1$ and $k_2$ integers (Fig.~\ref{fig:Fig3}d). 

For both the PSD and bispectrum, distinct pulse sequences have the effect of giving different weights to the comb teeth, granting access to complementary information about $S(k\omega_h)$ and $S_2(\vec k \omega_h)$, enabling their reconstruction. More specifically, in both cases the basic steps of our protocol consist of (i) applying a set of sufficiently distinct pulse sequences $p$ (Fig.~\ref{fig:Fig3}a); (ii) measuring the corresponding decay and phase parameters; and (iii) solving the resulting systems of linear equations, which give $\chi_p(MT)$ and $\varphi_p(MT)$ as a function of $S(k\omega_h)$ and $S_2(\vec k \omega_h)$. 
Since classical noise 
has a spectrum with even symmetry, $S(\omega)=S(-\omega)$, the PSD is specified 
across all frequency space 
by its values at positive frequencies. Likewise, the bispectrum is completely specified by its values over a subspace ${\mathcal D}_2$ known as the principal domain~\cite{Chandran1994,Norris2016}, illustrated in Fig.~\ref{fig:Fig3}d. Reconstructing the bispectrum over ${\cal D}_2$ and exploiting the symmetries that $S_2(\vec\omega)$ exhibits  (shown in Fig.~\ref{fig:Fig3}c) thus suffices to retrieve the bispectrum over the whole relevant frequency domain.

\begin{figure}[t!]
\centering
\includegraphics[width=8.9cm]{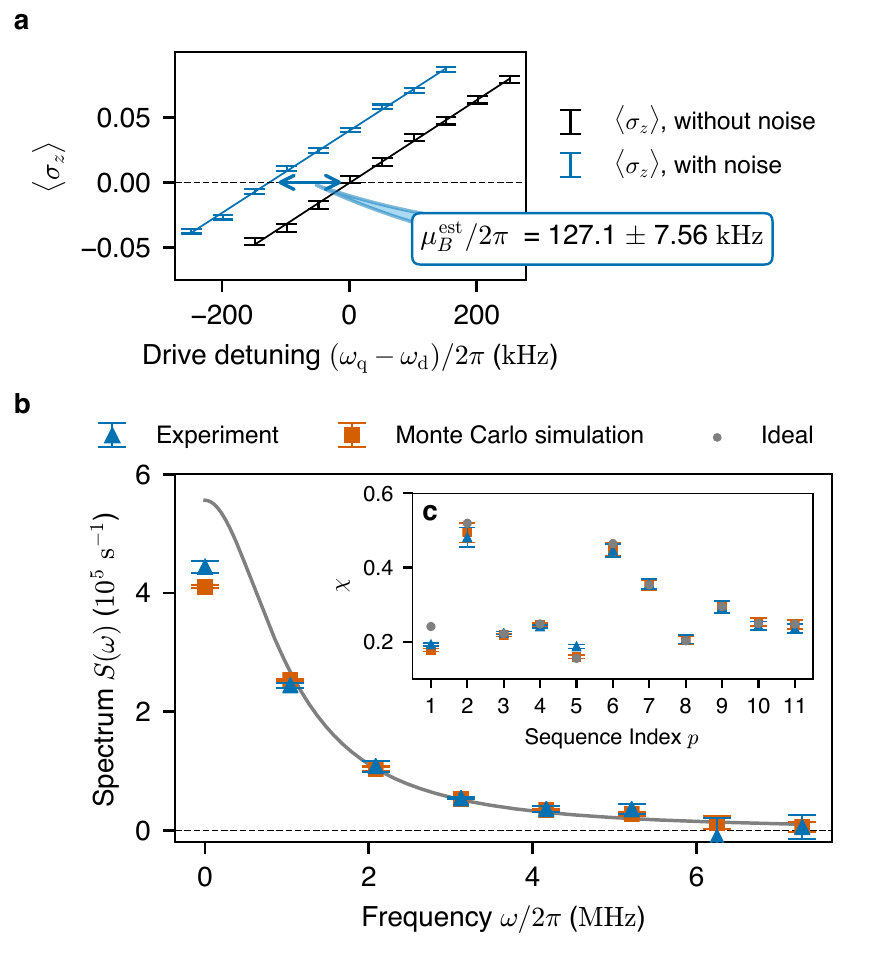}
\caption{\textbf{Gaussian spectral estimation: noise mean and PSD.}
\textbf{a}, Measured values of $\langle \sigma_z \rangle$ after a 50~ns-long Ramsey sequence vs.  drive detuning $D=\omega_\mathrm q-\omega_\mathrm d$. The separation between the $x$-intercepts of the two fitted lines gives the mean $\mu_B^{\mathrm{est}}$ of the injected dephasing noise. 
\textbf{b}, Comparison of the experimental reconstruction (blue triangle) and Monte Carlo simulation (orange square) with the ideal PSD (gray solid line). 
\textbf{c}, Decay constants $\chi$. Except for $p$ = 1, the ideal data (gray circles) are in very good agreement with both the experimental results and Monte Carlo simulations. Error bars represent 95\% confidence intervals.}
\label{fig:Fig4}
\end{figure}

Figure~\ref{fig:Fig4} presents experimental results for determining the mean and PSD, which suffice to characterize the noise process in the Gaussian approximation. To measure $\mu_B$ by Ramsey interferometry, we apply a pair of $\pi/2$ pulses with a drive at frequency $\omega_\mathrm d$, first about $x$ at time $t=0$ ($R_x(\pi/2)$), and then about $y$ at time $t=T$ ($R_y(\pi/2)$). We choose a pulse interval $T=50$ ns, which is short enough for cumulants of order higher than one to be negligible, but long enough to avoid pulse overlap. The qubit polarization at time $t_f$ after the two pulses is then $\langle \sigma_z(t_f)\rangle\approx(D+\mu_B)T'$, where $D\equiv\omega_\mathrm q-\omega_\mathrm d$ is the drive detuning, and $T'$ is an effective time interval that accounts for the finite-width pulse shape (see Supplementary  Note~
6). Thus, plotting $\langle \sigma_z(t_f)\rangle$ as a function of $D$ produces a straight line whose $x$-intercept is $-\mu_B$, leading to an estimate that is insensitive to the pulse shape to first order in the cumulant expansion. Figure~\ref{fig:Fig4}a presents data for measurements of $\langle \sigma_z(t_f)\rangle$, and shows how we isolate the contribution of the engineered noise source by performing the sequence with (blue data set) and without (black data set) applied noise. The mean of the engineered noise is estimated by subtracting the $x$-intercepts of the straight lines that are fitted to each data set. Performing these fits under the conditional normal model of linear regression
(see Supplementary Note~
6) yields 
the estimate $\mu_B^\mathrm{est}/2\pi=127.1\pm7.56$~kHz, where the uncertainty corresponds to the 95\% confidence interval calculated from the asymptotic normal distribution of qubit polarization.

To estimate the PSD by the comb approach outlined above, we use both a period of free evolution ($p=1$) and $M=10$ repetitions of base sequences $p=2,\ldots,11$ illustrated in Fig.~\ref{fig:Fig3}a (see Supplementary Note~
4 for the actual pulse times). For $M\gg1$, the FF entering the decay constant in Eq.~(\ref{eqnchi}) becomes approximately $|F_p(\omega,MT)|^2\approx\frac{M}{T}|F_p(\omega,T)|^2\sum_{k=-\infty}^\infty\delta(\omega-k\omega_h)$, which enables us to sample the PSD at the harmonic frequencies  in terms of the (known) control FFs,

\begin{align}
\chi_p(MT)\approx\frac{M}{T}\sum_{k\in \mathcal{K}_1}
|F_p(k\omega_h,T)|^2S(k\omega_h). 
\end{align}

Here, we have used the even symmetry of the PSD, and the high-frequency decay of the PSD and FFs to truncate the comb to a finite set of positive harmonics, ${\cal K}_1\equiv \{0,\ldots,K-1\}$. Rather than solving the above linear system by matrix inversion as in Ref. \cite{Alvarez2011}, we employ a statistically-motivated maximum likelihood estimate (MLE), which takes experimental error into account (see Supplementary  Note~
7). Using measurements of $\chi_p(MT)$ for each of the same $P=11$ control sequences to be used for the bispectrum estimation, we find a well-conditioned system for $K=8$. 

Figure~\ref{fig:Fig4}b compares the experimentally estimated PSD at the $K=8$ harmonics (blue triangles) with the ideal PSD obtained from Eq.~\eq{eqnS} for our engineered noise (solid gray line) and Monte Carlo simulations of the QNS protocol (orange squares). The experimental and simulated estimates of the PSD are plotted along with 95$\%$ confidence intervals obtained from the asymptotic normal distribution of the decay constants. Figure~\ref{fig:Fig4}c shows the experimental and simulated values of $\chi_p(MT)$ that were used as input for the reconstructions, along with ideal values obtained by substituting Eq.~\eq{eqnS} into Eq.~\eq{eqnchi} and approximating the FF by  the ideal (infinite) comb as given above. The PSD is slightly underestimated at zero frequency in both the experiment and Monte Carlo simulation since the FF of sequence $p=1$ (a 1 $\upmu \mathrm{s}$-long free induction decay) is comparable in bandwidth to the PSD, whereas the reconstruction procedure assumes  the PSD is sampled by infinitely narrow FFs. The disagreement of the experimental and simulated $\chi_p(MT)$ for $p=1$ with the ideal value is also explained by the non-negligible bandwidth of the FF (Fig.~$\ref{fig:Fig4}$c). Apart from these well-understood discrepancies at $\omega=0$, the quantitative agreement of the experimental reconstruction with simulations and ideal values is remarkable, which demonstrates that our protocol is able to reliably characterize Gaussian features of the applied noise.

\begin{figure*}
\centering
\includegraphics[width=\textwidth]{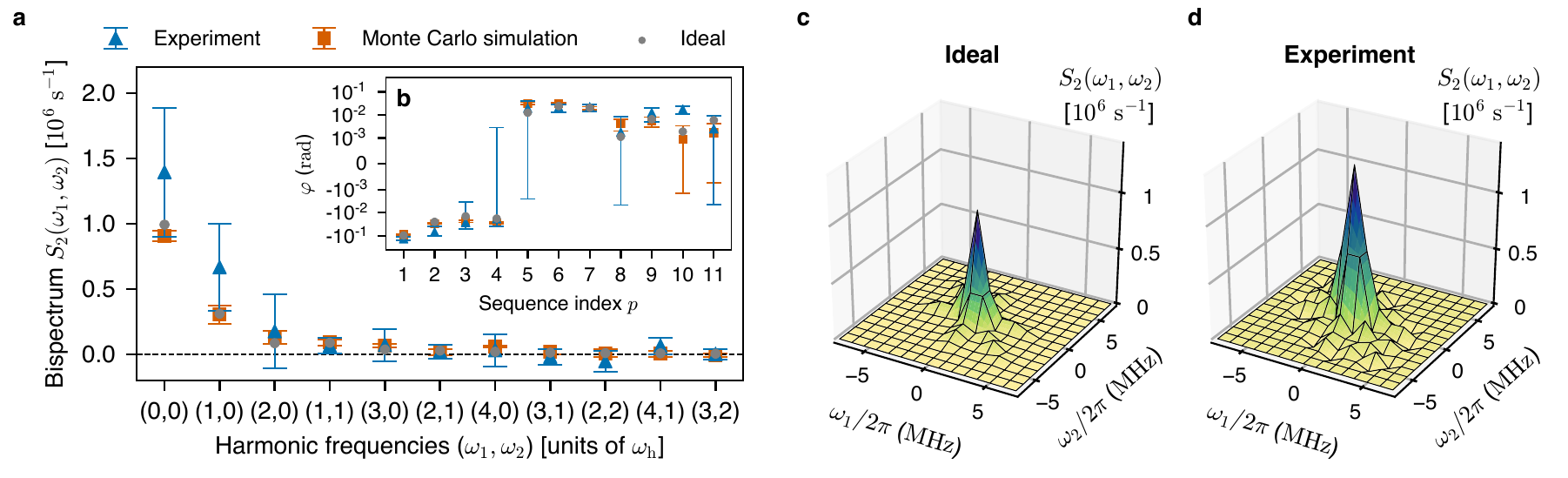}
\captionof{figure}{\textbf{Non-Gaussian spectral estimation: noise bispectrum.}
\textbf{a}, Experimental data (blue triangles), Monte Carlo simulations (orange squares) and ideal values (gray circles) for the bispectrum of the engineered dephasing noise. The error bars indicate that the experimental bispectrum agrees with both the ideal bispectrum and Monte Carlo simulations of the protocol within 95$\%$ confidence intervals. 
\textbf{b}, Estimated non-Gaussian phase angles $\varphi$. Error bars represent 95\% confidence intervals.
\textbf{c}, 3D visualization of the ideal bispectrum. 
\textbf{d}, 3D visualization of the reconstructed bispectrum for the experimental data. }
\label{fig:Fig5}
\end{figure*}

We are now in a position to present our key result: the reconstruction of the noise bispectrum. As anticipated, this entails a higher-dimensional analogue of the comb-based approach used for the PSD. We estimate the non-Gaussian phase given in Eq.~\eq{eqnphi} by subtracting the contribution of the noise mean from the total measured phase, $\varphi_p(MT)  = \phi_p(MT) - \mu_B F_p(0,MT) $, where we replace $\mu_B$ by $\mu_B^\mathrm{est}$ experimentally determined above. After $M\gg 1$ repetitions of sequence $p$, the FF becomes a 2D comb (Fig.~\ref{fig:Fig3}d), and the non-Gaussian phase becomes a sampling of the bispectrum at the harmonics $\vec k\omega_h$, that is, 
$\varphi_p(MT)=-\frac{M}{3!T^2}\sum_{\vec{k}\in\mathbb{Z}^2}G_p(\omega_h\vec{k},T)S_2(\omega_h\vec{k}\,).$ Since both the filter and bispectrum decay at high frequencies, we can truncate this sum to a finite number of $\vec{k}=(k_1,k_2)$. As the bispectrum 
is completely specified by its values on the principal domain, we may further restrict our consideration to a subset of harmonics, $\mathcal{K}_2\equiv \{\vec{k}_1,\ldots ,\vec{k}_N\} \subset {\mathcal D}_2$ (Fig. \ref{fig:Fig3}c). The non-Gaussian phase then becomes

\begin{align}
\varphi_p(MT)=-\frac{M}{3! T^2}\sum_{\vec{k}\in\mathcal{K}_2}m(\omega_h\vec{k})\,\text{Re}[G_p(\omega_h\vec{k},T)]S_2(\omega_h\vec{k}\,),
\end{align}

where the multiplicity $m(\omega_h\vec{k})$ accounts for the number of points equivalent to  $S_2(\omega_h\vec{k}\,)$ by the symmetry properties of the bispectrum. Also on account of these symmetries, the imaginary component of $G_p(\omega_h\vec{k},T)$ cancels when the sum is restricted to $\mathcal{D}_2$ (see Supplementary Note 
8).

By measuring the non-Gaussian phase for $P\geq N$ different control sequences, we can construct a vector $\vec{\varphi}=[\varphi_1(MT),\ldots,\varphi_P(MT)]^T$ and a linear system of the form  

\begin{align}
\vec{\varphi}=\mathbf{A}\vec{S}_2,\;\;\;\mathbf{A}_{pn}\!=-\frac{M}{3! T^2}\,m(\omega_h\vec{k}_n)\,\text{Re}[G_p(\omega_h\vec{k}_n,T)],
\end{align}

where $\vec{S}_2=[S_2(\omega_h\vec{k}_1),\ldots,S_2(\omega_h\vec{k}_N)]^T$ contains the bispectrum at the harmonics in $\mathcal{K}_2$ and  $\mathbf{A}$ is a $P\times N$ reconstruction matrix. The simplest way to estimate the bispectrum from this linear system is the least-squares estimate employed in Ref. \cite{Norris2016}, involving the (pseudo-)inverse of the reconstruction matrix, $\vec{S}^\mathrm{est}_2=\mathbf{A}^{-1}\vec{\varphi}$. As in the case of PSD estimation, a potential drawback of this inversion-based approach is numerical instability stemming from an ill-conditioned $\mathbf{A}$, which occurs when the FFs have a high degree of spectral overlap. Since ill-conditioning makes the least-squares estimate sensitive to even small errors in the measured phases, we again utilize a maximum-likelihood approach with optional regularization to further increase stability (see Supplementary Note 
8). From the asymptotic Gaussian distribution of the measurement outcomes of $\vec{\varphi}$, the regularized maximum-likelihood estimate (RMLE) is found as 
\begin{align}
\vec{S}_2^{\,\text{RMLE}} \!= \argmin_{\vec S_2} \!\left[\frac{1}{2} (\mathbf{A} \vec{S}_2 - \vec{\varphi})^T \mathbf{\Sigma}^{-1} (\mathbf{A} \vec{S}_2 - \vec{\varphi}) \! + \!
|\!| \lambda \mathbf{D}\vec{S}_2|\!|_2^2 \right]\!, 
\label{eqnRMLE}
\end{align}
where $\! |\!| \,\cdot\, \! |\!|_2$ denotes the $L_2$-norm and $\lambda \geq 0$ parametrizes the strength of the regularization ~\cite{Hansen2000}. 
Due to its dependence on the covariance matrix $\mathbf{\Sigma}$, the RMLE down-weights phase measurements with larger error. Numerical stability is increased by the regularizer $|\!|\lambda\mathbf{D}\vec{S}_2|\!|_2^2$, which acts as an effective constraint.  When the smoothing matrix 
$\mathbf{D}$ is proportional to $\mathbf{I}$, the regularizer reduces to the well-known Tikhonov (or $L_2$) form. Since the numerical stability afforded by regularization comes at the cost of additional bias, choosing the regularization strength is a nontrivial task. In Supplementary  Note 
8, we detail how we have selected $\lambda$ based on the so-called ``L-curve criterion''. Interestingly, since $\mathbf{A}$ is sufficiently well-conditioned for the sequences we have chosen, we find that regularization gives negligible benefit. Accordingly, we use $\lambda = 0$ (which recovers standard MLE) in our experimental reconstructions.

Figure~\ref{fig:Fig5}a compares the results of the non-Gaussian spectral estimation for the harmonics in the principal domain for the experiment (blue triangles) with both the ideal bispectrum obtained from Eq.~\eq{eqnS2} (gray circles) and from Monte Carlo simulations (orange squares). To estimate the experimental bispectrum, we input the measured data for $\vec{\varphi}$ and $\mathbf{\Sigma}$ shown in Fig.~\ref{fig:Fig5}b into $\vec{S}_2^{\,\text{RMLE}}$ given by Eq.~\eq{eqnRMLE}. The ideal values of $\varphi_p$, also shown in Fig.~\ref{fig:Fig5}b, are obtained by substituting Eq.~\eq{eqnS2} into Eq.~\eq{eqnphi}. We further display 3D representations of the full bispectra, obtained by applying relevant symmetries to the data on ${\cal D}_2$, for the ideal (Fig.~\ref{fig:Fig5}c) and experimental (Fig.~\ref{fig:Fig5}d) cases, respectively. Ignoring error bars, the reconstructed bispectrum appears to be an over-estimate with respect to the ideal one. This error may be attributed to 
noise during the finite-duration control pulses used in the experiment, leading to effective pulse infidelity. Upon taking the error bars in Fig.~\ref{fig:Fig5}a into consideration, however, the ideal and simulated values of the bispectrum lie within the 95$\%$ confidence intervals of the experimental reconstruction, suggesting that this estimation error is statistically insignificant and thus successfully extending the validation of our QNS protocol to the leading non-Gaussian noise cumulant.

Although the theoretical bispectrum falls within the 95$\%$ confidence interval of the estimate, reducing the magnitude of uncertainties is clearly necessary to push the application of non-Gaussian QNS to uncontrollable native noise, whose strength may be comparatively weak. We note that the spectral characterization of the non-Gaussian noise process engineered in this experiment requires an extremely precise estimation of $\mu_B$. Since reconstructions of the bispectrum are obtained using $\varphi_p(MT)=\phi_p(MT)-\mu_B F_p(0,MT)$, the uncertainty in $\mu_B^\mathrm{est}$ propagates to $\varphi_p(MT)$ when $p$ has zero filter order, i.e. $F_p(0,MT)\neq 0$. These sequences play a crucial role in estimating the bispectrum at the ``zero points", grid points $(\omega_1,\omega_2)$ with $\omega_1=0$ or $\omega_2=0$. Since $\mu_B$ is much larger than the third cumulant for the current noise process, even a small relative uncertainty in $\mu_B^{\mathrm{est}}$ can lead to greater error  in the bispectrum estimate at the zero points, as the error bars in Fig.~\ref{fig:Fig5}a attest.

\section*{Discussion}

In summary, we reported the first experimental demonstration of high-order spectral estimation in a quantum system. By producing and sensing engineered noise with well-controlled cumulants, we were able to successfully validate a spectroscopy protocol that reconstructs both the power spectral density and the bispectrum of non-Gaussian dephasing noise. Our theory and experimental demonstration lay the groundwork for future research aiming at complete spectral characterization of realistic non-Gaussian noise environments in quantum devices and materials. Theoretically, we expect that the regularized maximum-likelihood estimation approach to quantum noise spectroscopy we invoked here will prove crucial to ensure stable spectral reconstructions in more general settings. Devising alternative estimation protocols based on optimally band-limited control modulation and multitaper techniques \cite{Birkelund} appears especially compelling, in view of recent advances in the Gaussian regime \cite{Frey2017,Norris2018}. We believe that obtaining a complete spectral characterization will ultimately provide deeper insight into the physics and interplay of different microscopic noise mechanisms, including non-classical non-Gaussian noise, as possibly arising from photon-number mediated non-linear couplings \cite{Fei2018}. 

\smallskip

\section*{Data availability}
The data that support the findings of this study may be made available from the corresponding authors upon request and with the permission of the US Government sponsors who funded the work.

\section*{Code availability}
The code used for the analyses may be made available from the corresponding authors upon request and with the permission of the US Government sponsors who funded the work.

\printbibliography

\section*{Acknowledgement}
It is a pleasure to thank K. Harrabi, M. Kjaergaard, P. Krantz, G. A. Paz-Silva,  J.I.J. Wang, and R. Winik for insightful discussions, and M. Pulido for generous assistance. We thank H. Bethany for the SEM image of the device. This research was funded by the U.S. Army Research Office grant No. W911NF-14-1-0682 (to L.V. and W.D.O.); and by the Department of Defense via MIT Lincoln Laboratory under Air Force Contract No. FA8721-05-C-0002 (to W.D.O.). Y.S. and F. B. acknowledge support from the Korea Foundation for Advanced Studies and from the Fonds de Recherche du Qu\'ebec -- Nature et Technologies, respectively. The views and conclusions contained herein are those of the authors and should not be interpreted as necessarily representing the official policies or endorsements, either expressed or implied, of the U.S. Government.

\section*{Author contribution}
Y.S., F.Y., and S.G. performed the experiments. F.B. and Y.S. carried out numerical simulations and analyzed the data and L.V., L.M.N., S.G. and W.D.O. provided feedback. L.M.N., F.B., and L.V. designed the pulse sequences and developed the estimation protocol used in the experiment. F.Y. and S.G. designed the device and D.K.K. and J.L.Y. fabricated it. J.Y.Q. and U.L. provided experimental assistance. Y.S., F.B., L.M.N., and L.V. wrote the manuscript with feedback from all authors. L.V., S.G., T.P.O., and W.D.O. supervised the project.

\section*{Competing Interests}
The authors declare no competing interests.
\clearpage 
\newpage 

\onecolumn

\section*{Supplementary Information}
\renewcommand{\thefigure}{S\arabic{figure}}
\setcounter{figure}{0}
\renewcommand{\thetable}{S\arabic{table}}
\setcounter{table}{0}


\section{Additional experimental and computational detail}

\subsection{Device parameters and fabrication of the qubit sensor}
\label{sec:device_params}

Device parameters are summarized in Table~\ref{table:device_params}.

\begin{table}[h!]
\caption{\textbf{Device parameters.}} 
\centering 
\begin{tabular}{c | c} 
\hline \hline 
Parameter &  Value \\ \hline 
Qubit frequency $\omega_0/2\pi$& 2.920 GHz  \\ 
Qubit anharmonicity $A/2\pi$& 1.163 GHz  \\ 
Relaxation time $T_1$ & 27.0 $\pm$ 2.7 $\upmu \mathrm{s}$ \\ 
Spin-echo relaxation time $T_2$ & 35.9 $\pm$ 4.4 $\upmu \mathrm{s}$ \\ 
Free induction decay time $T_2^*$ & 12.2 $\pm$ 1.0 $\upmu \mathrm{s}$ 
\\ 
Readout cavity frequency $\omega_r/2\pi$ & 7.348 GHz \\ 
Readout cavity linewidth $\kappa/2\pi$ & 2.548 MHz \\ 
Dispersive coupling strength $\chi/2\pi$ & 0.130 MHz \\ \hline \hline
\end{tabular}
\label{table:device_params}
\end{table}

The device was fabricated in the same way as in Ref.~\cite{Yan2016}. It is a generalized version of the capacitively shunted flux qubit~\cite{Yan2016}, comprising a capacitively shunted small-area junction in parallel is a series array of $N$ junctions. In the capacitively shunted flux qubit of Ref.~\cite{Yan2016}, this array comprised $N=2$ junctions.  Here, the number of array junctions is $N=8$, far fewer than used in the fluxonium regime of operation [S1]. The area of each Josephson junction forming the array is identical and designed to be $0.2 \times 1.2~\upmu \textrm{m}^2$. The left junction in Fig.~\ref{fig:Fig1}a is smaller in area by a factor of 8 than the right junction ($\alpha = 1/8$). The critical current density $J_{\mathrm{c}}$ is measured to be 
$0.60\pm 0.01~\upmu \mathrm{A} / \upmu \mathrm{m}^2$ and the shunt capacitance $C_{\mathrm{sh}}$ is designed to be 20~$\mathrm{fF}$.


\subsection{Randomized benchmarking of single-qubit gates}
\label{sec:1QB_RB}

\begin{figure}[h!]
\centering
\includegraphics{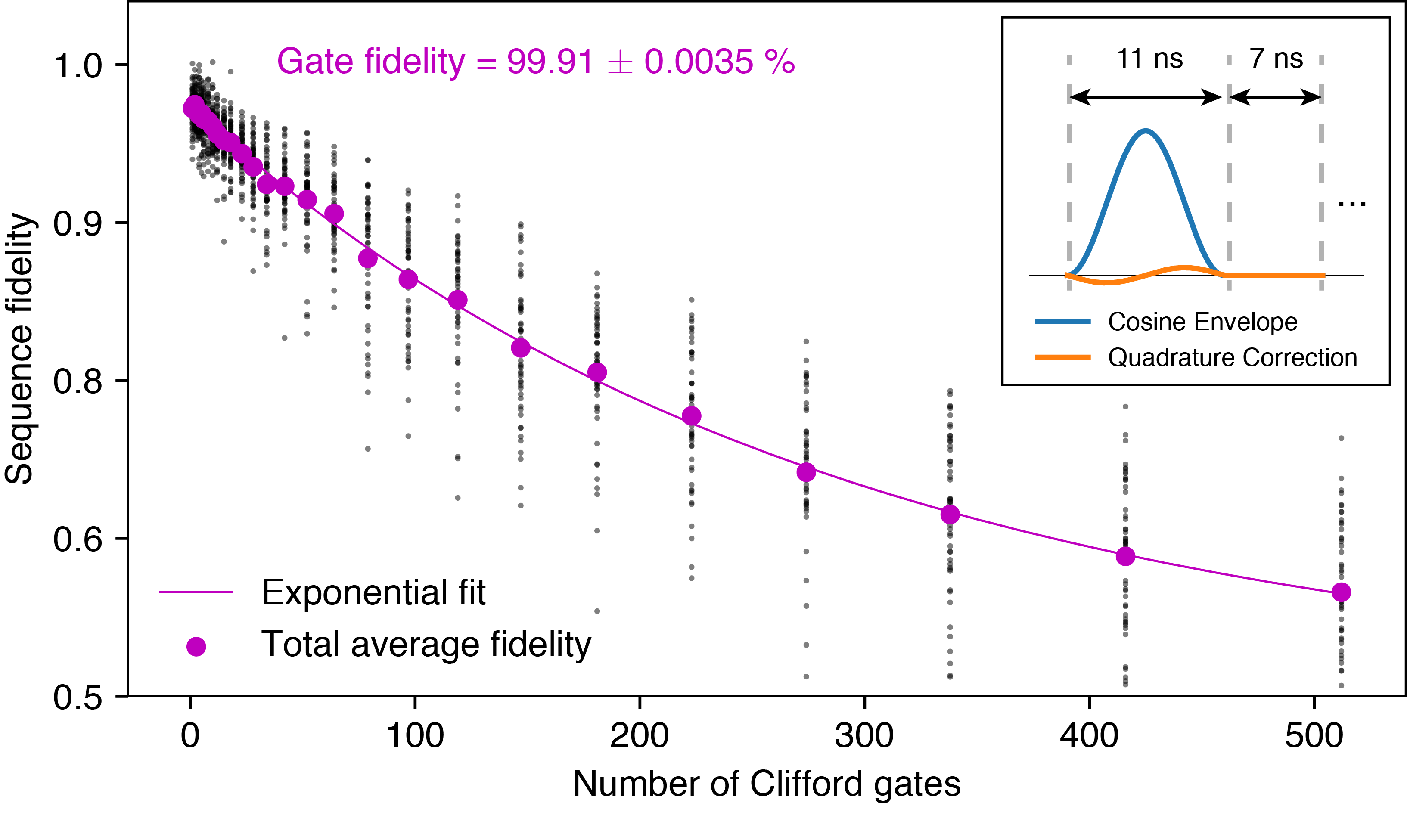}
\caption{\textbf{Randomized benchmarking of single-qubit gates.} Standard single-qubit Clifford randomized benchmarking, indicated as average sequence fidelity (magenta circle) vs. number of Clifford gates. There are 50 randomizations for each number of Clifford gates. The envelope of microwave pulse is a cosine with a total length of 11~ns; a constant buffer time of 7~ns is inserted after each pulse to ensure complete separation of the pulses (inset).} 
\label{fig:SuppFig1}
\end{figure}

We characterized an average error rate of single-qubit gates by performing Clifford randomized benchmarking [S2]
(Fig.~\ref{fig:SuppFig1}). As mentioned in the main text, single-qubit operations are performed using cosine-shaped microwave pulses, applying 
a quadrature correction (DRAG~\cite{Motzoi2009}) to minimize unwanted phase evolution and leakage due to the presence of higher levels. 

\subsection{Measurement setup}

\subsubsection{Cryogenic setup}

The experiments were performed using a Leiden CF-450 dilution refrigerator, capable of cooling to a base temperature of 15 mK. The samples were magnetically shielded with a superconducting can surrounded by a Cryoperm-10 cylinder. The schematic of the cryogenic circuitry is shown in Figure \ref{fig:SuppFig2}. There are two RF lines for the input and the output of the samples; applying microwave readout tone and measuring the transmission of sample respectively. Thermal noise from room temperature on the RF drive lines is attenuated with 20 dB at the 3 K stage, followed by 6 dB at still, and 26 dB at the 20 mK stage. All attenuators in the cryogenic samples are made by XMA. Note that there is one additional RF line for pumping the Josephson traveling wave parametric amplifier (JTWPA) [S3]
used as a first-stage pre-amplifier to amplify the readout signal at base temperature. The effective noise temperature is determined primarily by the JTWPA, with a total system noise measured to be about 600 mK. To avoid any back-action of the pump-signal from TWPA, we added a microwave isolator between the  samples and the TWPA. On the RF output line, there is a high-electron mobility transistor (HEMT) amplifier (Cryo-1-12 SN508D) at the 3 K stage. Two microwave isolators allow for the signal to pass through to the amplifier without being attenuated, while taking all the reflected noise off of the amplifier and dumping it in a 50 $\Omega$ termination instead of reaching the sample. 

There are two additional lines for qubit flux bias: one is for DC flux bias, which is applied globally through a coil installed in the device package, the other is to apply magnetic flux to the qubit thorough a local antenna. The primary requirement of the DC flux bias line is the ability to tune through at least a single flux quantum on the SQUID of the qubit with high precision and low noise. The local flux bias line is attenuated by 20 dB at the 3 K stage, 6 dB at the still, 20 dB at the 20 mK stage to remove excess thermal photons from higher-temperature stages.  

\subsubsection{Room temperature control}

\begin{figure}[h!]
\centering
\includegraphics{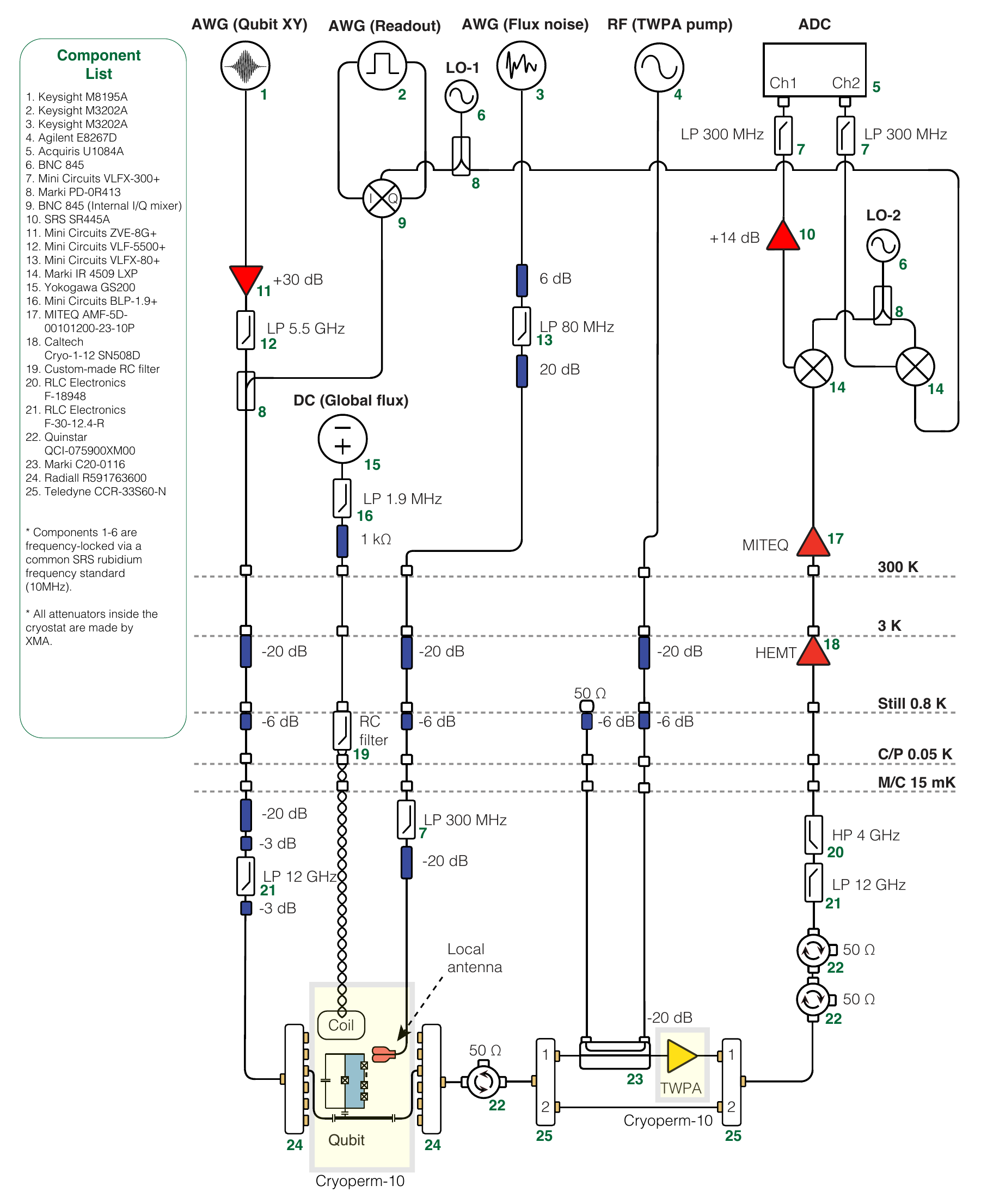}
\caption{\textbf{Electronics and control wiring}}
\label{fig:SuppFig2}
\end{figure}

Outside of the cryostat, we have all of the control components which allow us to apply microwave signals that address the cavity and the qubits, as well as the components necessary to resolve the readout signal. All the signals are added using microwave power splitters (Marki PD0R413) used in reverse. Direct digital synthesis of the qubit signals is performed using a high-speed arbitrary waveform generator (AWG Keysight M8195A), which has a 65 GS/sec sampling rate and sufficient bandwidth for this purpose. The output line is further amplified outside of the cryostat with an amplifier (MITEQ AMF-5D-00101200-23-10P) with a quoted noise figure of 2.3 dB, and a preamplifier (Stanford Research SR445A). A detailed schematic is given in Fig~\ref{fig:SuppFig2}. We use an IQ demodulation technique 
to mix down the signal entering the RF port with a reference signal detuned by 40 MHz applied to the LO port. This results in down-converted signals to 40 MHz using a mixer. All components are frequency-locked via a common SRS rubidium frequency standard (10 MHz).

\subsubsection{Pulse generation}
Qubit control pulse generation is performed via a Keysight M8195A AWG. The pulses are programmed in Labber and then uploaded to the Keysight M8195A. 

\clearpage
\newpage

\subsection{Generation of engineered noise} 
\label{secNoiseGeneration}
 
To synthesize a zero-mean flux-noise process $\Delta\Phi(t)\equiv\Phi(t)-\Phi_0/2$ with a given PSD, we use an AWG to produce sample waveforms consisting of $N_h$ harmonics
\begin{equation}
\Delta\Phi(t)=\sum_{m=1}^{N_h}\left(a_m\cos\omega_mt+b_m\sin\omega_mt\right),
\end{equation}
where $\omega_m=2\pi m/T_0$, and with $2\pi/T_0$ the fundamental angular frequency. The Fourier coefficients $a_m$ and $b_m$ are random variables with
\begin{align}
\mathbb E[a_m]=\mathbb E[b_m]=0,\quad\forall\,m,\hspace{5mm}
\mathbb E[a_m b_n]=0, \quad \forall m,n,\hspace{5mm}
\mathbb E[a_m a_n]=\mathbb E[b_m b_n]=0,\quad\forall\,m\neq n.
\end{align}
Further taking $a_m$ and $b_m$ to have normal (Gaussian) distributions with variance $\sigma_m^2=2 S_\Phi(\omega_m)/T_0$, the waveforms $\Delta\Phi(t)$ become a discrete approximation of a Gaussian stochastic process with a frequency-domain PSD $S_\Phi(\omega)$.

In all experiments presented in the main text, we consider
\begin{align}
S_\Phi(\omega)=\frac{P_0/\pi\omega_c}{1+(\omega/\omega_c)^2}.
\end{align}
with $P_0$ the noise power and $\omega_c/2\pi=0.5$ MHz. For the experiment presented in Fig.~\ref{fig:Fig2} of the main text, to produce a discrete approximation of a noise process with this spectrum, we take $T_0=20$~$\upmu$s and $N_h=10^3$, corresponding to harmonics separated by the fundamental frequency $1/T_0=50$~kHz with a high-frequency cutoff at $\omega_{N_h}/2\pi=N_h/T_0=50$~MHz. For the experiment presented in Figs.~\ref{fig:Fig4} and \ref{fig:Fig5} of the main text, we take $T_0=200$~$\upmu$s and $N_h=10^4$, yielding harmonics separated by $1/T_0=5$~kHz with the same high-frequency cutoff at $\omega_{N_h}/2\pi=N_h/T_0=50$~MHz. A new waveform is produced by the AWG for each measurement of a Pauli operator performed on the qubit to ensure statistical independence of the samples of the stochastic process, leading to a total number of noise samples of 40,000 for the experiment presented in Fig.~\ref{fig:Fig2}, and 80,000 for Figs~\ref{fig:Fig4} and \ref{fig:Fig5}. The waveforms have a duration of $1.25$~$\upmu$s for the experiment presented in Fig.~\ref{fig:Fig2} and $20$~$\upmu$s for Figs.~\ref{fig:Fig4} and \ref{fig:Fig5}.

\subsection{Control pulse sequences} 
\label{sec:controlpulse}

\begin{figure}[h!]
 \centering
 \includegraphics{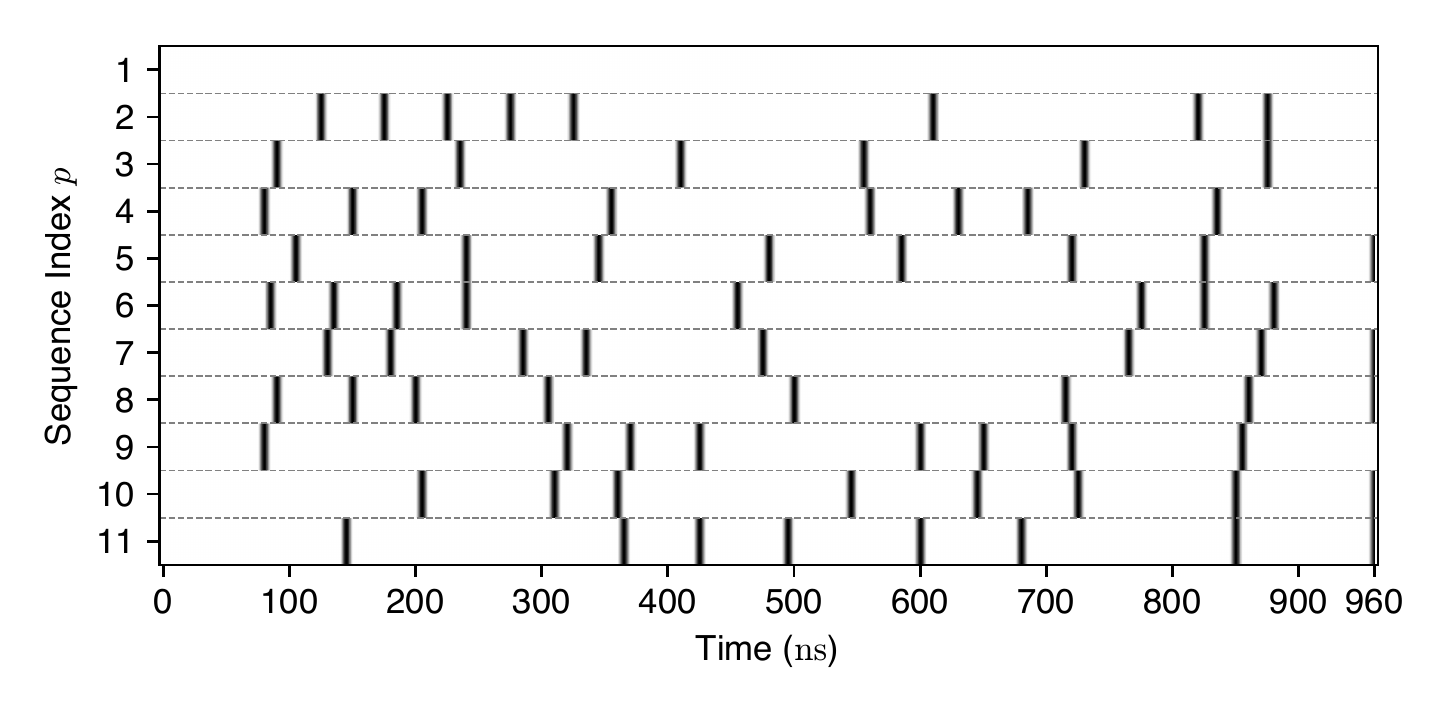}
 \caption{\textbf{Timing diagrams of the base control sequences.} Only $\pi$ pulses are shown.}
\label{fig:SuppFig3}
\end{figure}

The set of control pulse sequences designed for reconstructing the bispectrum are summarized in Table~\ref{table:seq_DD} and visualized in Fig.~\ref{fig:SuppFig3}. Note that all control pulse sequences start and end with a $\pi/2$ pulse for the purposes of state preparation and tomography.
 
\begin{table}[h!]
\caption{\textbf{Control pulse sequences designed for non-Gaussian spectral estimation.} } 
\centering 
\begin{tabular}{c | c | c | c} 
\hline \hline 
Seq. Index $p$ & Position of $\pi$ pulses [$\mathrm{ns}$] & Repetitions $M$ & Filter function at zero frequency $F_p(0,T)$   \\ \hline 
1 & No pulses (free evolution) & 1 & $\neq$ 0  
\\ \hline
2 & 125, 175, 225, 275, 325, 610, 820, 875 & 10 & $\neq$ 0
\\ \hline
3 & 90, 235, 410, 555, 730, 875 & 10 & $\neq$ 0 
\\ \hline
4 & 80, 150, 205, 355, 560, 630, 685, 835 & 10 & $\neq$ 0 
\\ \hline
5 & 105, 240, 345, 480, 585, 720, 825, 960 & 10 & $\neq$ 0 
\\ \hline
6 & 85, 135, 185, 240, 455, 775, 825, 880 & 10 & 0 
\\ \hline
7 & 130, 180, 285, 335, 475, 765, 870, 960 & 10 & 0 
\\ \hline
8 & 90, 150, 200, 305, 500, 715, 860, 960 & 10 & 0 
\\ \hline
9 & 80, 320, 370, 425, 600, 650, 720, 855 & 10 & 0 
\\ \hline
10 & 205, 310, 360, 545, 645, 725, 850, 960 & 10 & 0 
\\ \hline
11 & 145, 365, 425, 495, 600, 680, 850, 960 & 10 & 0 
\\ \hline \hline 
\end{tabular}
\label{table:seq_DD}
\end{table}

\subsection{Monte Carlo simulations 
\label{sec:monteCarlo}}

For the Monte Carlo simulations that are presented in the main text, we consider a single qubit controlled via a microwave drive at angular frequency $\omega_\mathrm d$, which is used to apply pulses about $\sigma_x$ and $\sigma_y$. In contrast with 
the main text, here we do not assume that these pulses are instantaneous. To describe the time-evolution of the qubit under the combined action of these finite-width pulses and classical noise described by the process $B(t)$, we consider the Hamiltonian in the lab frame,
\begin{align}
 H(t)=\frac{\omega_\mathrm q+B(t)}2\sigma_z + \varepsilon(t)\cos[\omega_\mathrm d t+\theta(t)]\sigma_x,
\end{align}
where $\omega_\mathrm q$ is the qubit angular frequency, and $\varepsilon(t)$ and $\theta(t)$ are the drive amplitude and phase, respectively.

We next move to the frame that rotates at the drive frequency by applying the unitary transformation $R_\mathrm d(t)=\exp(-i\omega_\mathrm d t\,\sigma_z/2)$, leading to the Hamiltonian $H_\mathrm d(t)=R^\dag_\mathrm d(t)H(t)R_\mathrm d(t)-iR^\dag_\mathrm d(t)\dot R_\mathrm d(t)$ in the rotating frame. This gives
\begin{align}
H_\mathrm d(t)=\frac{D+B(t)}2\sigma_z + H_\mathrm c(t),   
\label{eqnHd}
\end{align}
where $D\equiv\omega_\mathrm q-\omega_\mathrm d$ is the drive detuning and $H_\mathrm c(t)\equiv \varepsilon(t)\cos[\omega_\mathrm d t+\theta(t)][\sigma_+\exp(i\omega_\mathrm d t)+\mathrm{H.c.}]$ is the control Hamiltonian. Assuming that $\varepsilon(t)\ll\omega_\mathrm d$ and that $\theta(t)$ varies on a timescale much longer than $2\pi/\omega_\mathrm d$ allows us to invoke the rotating-wave approximation, under which terms oscillating like $\exp\{\pm[2i\omega_\mathrm d t+\theta(t)]\}$ are neglected. The resulting control Hamiltonian may be simplified as 
\begin{align}
H_\mathrm c(t)\approx\frac12\left[\eta_\mathrm I(t)\sigma_x+\eta_\mathrm Q(t)\sigma_y\right],   
\label{eqnHc}
\end{align}
where $\eta_\mathrm I(t)\equiv \varepsilon(t)\cos\theta(t)$ and $\eta_\mathrm Q(t)\equiv \varepsilon(t)\sin\theta(t)$ are the envelopes of the in-phase and quadrature components of the microwave control signal, respectively. Equation~\eq{eqnHc} then describes finite-width pulses about $x$ or $y$ axes.

We perform Monte Carlo simulations by solving the time-dependent Schr\"odinger equation associated with the Hamiltonian $H_\mathrm d(t)$ given by Eq.~\eq{eqnHd}, under the control Hamiltonian of Eq.~\eq{eqnHc}. The drive detuning $D$ is set to zero in the simulations. The envelope of each control pulse is cosine with total pulse duration 11 $\mathrm{ns}$ (see inset of Fig.~\ref{fig:SuppFig1}).

In our Monte Carlo simulations, we account for non-Gaussian noise by setting $B(t)\equiv \beta_\Phi \Delta\Phi(t)^2$ in Eq.~\eq{eqnHd}, and producing random samples of Gaussian flux-noise $\Delta\Phi(t)$ through the approach described in Section~\ref{secNoiseGeneration} of this Supplement. Because this approach relies on an exact solution of the qubit evolution under the noise samples, it is equivalent to accounting for all the terms in the cumulant expansion. The fundamental frequency $1/T_0$ and the number of harmonics $N_h$ involved in the Fourier-series representation of the noise process are the same as in Section~\ref{secNoiseGeneration}. To perform the simulations, we generate 100,000 noise samples for the data presented in Fig.~\ref{fig:Fig2} of the main text, and 80,000 noise samples for Figs.~\ref{fig:Fig4} and \ref{fig:Fig5}.

\section{Additional theoretical detail on estimation procedure}

\subsection{Estimation of the noise mean   
\label{sec:meanEstimate}}

\subsubsection{Ramsey estimation protocol}

\begin{figure}
\centering
\includegraphics{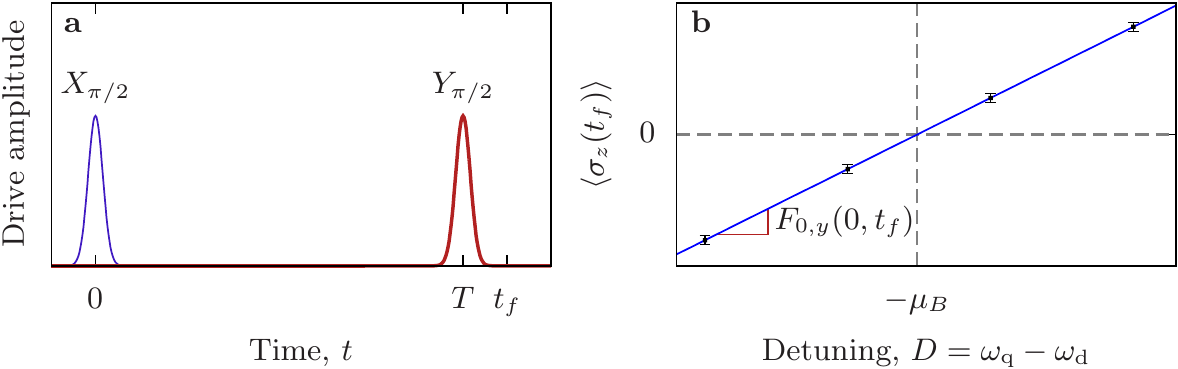}
\caption{Ramsey protocol for estimation of the noise mean. \textbf{a}, Envelopes $\eta_\mathrm I(t)$ (thin blue line) and $\eta_\mathrm Q(t)$ (thick red line) of the in-phase and quadrature components of the $\pi/2$ pulses applied about $\sigma_x$ and $\sigma_y$, respectively (see Eq.~\eq{eqnHc} in the text). \textbf{b}, Illustration of the technique for estimation of the mean by linear regression. Black error bars: Monte Carlo simulations of the $z$ component of the qubit polarization $\langle\sigma_z(t_f)\rangle$ after the pulse sequence as a function of the detuning of the drive from the qubit frequency. Blue line: linear regression. According to Eq.~\eq{eqnMeanEstimation}, the $x$-intercept of the blue line gives $-\mu_B$, and its slope gives the filter function $F_{0,y}(0,t_f)$, defined below Eq.~\eq{eqn:szgeneral}.}
    \label{fig:suppFigMean}
\end{figure}

To measure the noise mean $\mu_B$, we use the Ramsey sequence illustrated in Fig.~\ref{fig:suppFigMean}a. In this sequence, $\pi/2$ pulses about $\sigma_x$ and $\sigma_y$ are applied at times $t=0$ and $t=T$, respectively, followed by a measurement of the qubit in the $\sigma_z$ eigenbasis at time $t_f=T+\Delta T$, where $\Delta T$ is a buffer time much shorter than $T$, but longer than the pulse width. To lay down the theoretical basis of the procedure, we start from the rotating-frame Hamiltonian $H_\mathrm d(t)$ introduced in Eq.~\eq{eqnHd}, above. To describe the effects of control with finite-width pulses, it is useful to move to the toggling frame using the unitary transformation $R_\mathrm T(t)=\mathcal T\exp[-i\int_0^t ds\,H_\mathrm c(s)]$, with $\mathcal T$ the time-ordering operator, and where $H_c(t)$ is given by Eq.~\eq{eqnHc}. In this toggling frame, the Hamiltonian is
\begin{align}
H_\mathrm T(t)=\frac{D+B(t)}2\vec y_p(t)\cdot\vec\sigma,
\hspace{1.5cm}\vec{\sigma}\equiv(\sigma_x,\sigma_y,\sigma_z),
\end{align}
where $\vec y_p(t)$ has components $y_{p,a}(t)\equiv\frac12\mathrm{Tr}[R_\mathrm T^\dag(t)\sigma_z R_\mathrm T(t)\sigma_a]$, with $a\,\in\,\{x,y,z\}$, for pulse sequence $p$. Remark that a pulse sequence consisting of instantaneous $\pi$ pulses (instead of the Ramsey sequence considered here) would result in $\vec y_p(t)=[0,0,y_{p}(t)]$, where $y_{p}(t)$ is the switching function used in the main text.

Moving back to the lab frame, the expectation of the $z$ component of the qubit polarization after the pulse sequence is
\begin{align}
\langle \sigma_z(t_f)\rangle=\mathbb E\left\{\mathrm{Tr}\left[R_\mathrm T^\dag(t_f)\sigma_z R_\mathrm T(t_f)U_\mathrm T(t_f)\rho_0 U_\mathrm T^\dag(t_f)\right]\right\},
\label{eqn:szE}
\end{align}
where $U_\mathrm T(t)\equiv\mathcal T\exp[-i\int_0^tds\,H_\mathrm T(s)]$ is the time-evolution operator in the toggling frame, and $\rho_0$ is the initial qubit density matrix, before application of the pulses. We evaluate $\langle \sigma_z(t_f)\rangle$ perturbatively by performing a Dyson expansion of $U_\mathrm T(t)$. Assuming that $t_f$ is sufficiently short and that $D+B(t)$ is sufficiently small, we truncate this expansion to the first order in $D+B(t)$. Upon substitution into Eq.~\eq{eqn:szE}, this approach is equivalent to neglecting any contribution of cumulants of the noise beyond order 1. Taking $\rho_0\equiv|0\rangle\langle0|$, with $|0\rangle$ the eigenstate of $\sigma_z$ with eigenvalue $-1$, then yields
\begin{align}
\langle \sigma_z(t_f)\rangle \approx -y_{p,z}(t_f)+(D+\mu_B)\left[F_{p,x}(0,t_f)y_{p,y}(t_f)-F_{p,y}(0,t_f)y_{p,x}(t_f)\right],  \label{eqn:szgeneral}
\end{align}
where we have introduced the filter functions $F_{p,a}(\omega,t)\equiv\int_0^tds\,\mathrm e^{-i\omega s}y_{p,a}(s)$, with $a\,\in\,\{x,y,z\}$. For the pulse sequence illustrated in Fig.~\ref{fig:suppFigMean}a (which we label by $p=0$), neglecting any overlap between the pulses, it is straightforward to show that $\vec y_p(t_f)\equiv \vec y_0(t_f)=(-1,0,0)$ for $t_f=T+\Delta t$, leading to
\begin{align}
\langle \sigma_z(t_f)\rangle\approx(D+\mu_B)T',	
\label{eqnMeanEstimation}
\end{align}
where $T'\equiv F_{0,y}(0,t_f)=\int_0^{t_f}ds\,y_{0,y}(s)$ can be viewed as an effective pulse interval accounting for the shape of the pulses. For instantaneous pulses, $T'=T$.

According to Eq.~\eq{eqnMeanEstimation}, setting $D=0$, the first noise cumulant  $C^{(1)}(0)\equiv\mu_B$ may be estimated simply by measuring $\langle \sigma_z(t_f)\rangle $ and evaluating $\langle \sigma_z(t_f)\rangle/F_{0,y}(0,t_f)$. However, this approach requires accurate knowledge of $F_{0,y}(0,t_f)$, and thus of the shape of the control pulses. For the short Ramsey sequences required here to neglect cumulants of order higher than one in Eqs.~\eq{eqn:szgeneral}-\eq{eqnMeanEstimation}, the estimate of $\mu_B$ then becomes excessively sensitive to distortions of the pulse envelopes that occur in practice. As a result, estimates of $\mu_B$ become significantly biased. Since the bispectrum estimation technique that will be discussed in Sec.~\ref{sec:estimation:procedure} requires precise knowledge of $\mu_B$, this bias precludes accurate non-Gaussian QNS.

\subsubsection{Robust implementation via linear regression}

Crucially, the vulnerability of the above Ramsey scheme to pulse-width effects can be alleviated by estimating $\mu_B$ with a linear regression procedure. Indeed, according to Eq.~\eq{eqnMeanEstimation}, plotting $\langle \sigma_z(t_f)\rangle$ as a function of the detuning $D$ results in a straight line that intersects with the abscissa at $D=-\mu_B$ (Fig.~\ref{fig:suppFigMean}b), \emph{irrespective of} $F_{0,y}(0,t_f)$. Therefore, measuring $\langle \sigma_z(t_f)\rangle$ as a function of $D$ and performing a linear fit of the resulting data leads to an estimate of $\mu_B$ that is insensitive to the pulse shape to the first order in $D+B(t)$.

To apply this idea to our experimental data, we now explicitly construct an estimator of $\mu_B$ based on linear regression. For each drive detuning $D_j$, with $j\,\in\,\{1,2,\ldots,N_D\}$, we consider $N$ projective measurements of $\sigma_z$ yielding outcomes $Z_{j,i}=+1$ or $Z_{j,i}=-1$, where $i$ labels measurements. In the limit $N\gg1$, the sample mean $\overline Z_j$ of the projective measurements for detuning $D_j$ becomes Gaussian distributed,
\begin{align}
\overline Z_j\equiv\frac1N\sum_{i=1}^N Z_{j,i}
\sim \mathrm{Normal}\left[(D_j+\mu_B)F_{0,y}(0,t_f), \;\frac{\mathrm{var}(\sigma_z)_j}N\right], 
\label{eqn:conditionalNormal} 
\end{align}
where $\mathrm{var}(\sigma_z)_j\equiv\langle\sigma_z(t_f)^2\rangle_j-\langle\sigma_z(t_f)\rangle^2_j$ is the expected variance of $\sigma_z$ averaged over realizations of the noise process, for detuning $D_j$. Assuming that $\var(\overline Z_j)=\var(\sigma_z)_j/N$ is the same for all relevant detunings, $\var(\overline Z)_j\equiv\var(\overline Z)\,\forall\,j$, Eq.~\eq{eqn:conditionalNormal} then corresponds to the {\em conditional normal model of linear regression} [S4],
in which deviations of the data points from the expected linear behavior are given by independent and identically distributed (i.i.d.) Gaussian random variables. This assumption of uniform variance is justified in an approximate sense for ideal projective measurements. Indeed, in this situation, $\var(\overline Z_j)=[1-\langle \sigma_z(t_f)\rangle_j^2]/N$, so that $\var(\overline Z_j)$ is independent of $D_j$ to first order in $\langle \sigma_z(t_f)\rangle_j\approx(D+\mu_B)F_{0,y}(0,t_f)$, with $\var(\overline Z_j)\approx1/N$. Limiting ourselves to detunings for which $\langle \sigma_z(t_f)\rangle_j \lesssim 0.05$ (see Fig.~\ref{fig:Fig4}a), we find that $\var(\overline Z_j)$ (estimated from the sample mean of measurements of $\sigma_z$) varies by less than $5\%$ across values of $D_j$.

To define our estimator of $\mu_B$ within the conditional normal model of linear regression, we first introduce the quantities $a\equiv \mu_B F_{0,y}(0,t_f)$ and $b\equiv F_{0,y}(0,t_f)$, corresponding to the $y$-intercept and slope of the linear equation $\langle \sigma_z(t_f)\rangle_j=(D_j+\mu_B)F_{0,y}(0,t_f)\equiv a+b\,D_j$, respectively. Maximizing the likelihood of $a$ and $b$ with respect to measurement outcomes $\overline Z_j$ with the probability distribution given by Eq.~\eq{eqn:conditionalNormal} then yields the estimators
\begin{align}
    a^\mathrm{est} &= N_D^{-1}\sum_{j=1}^{N_D}\left(\overline Z_j-b^\mathrm{est} D_j\right),  \label{eqn:a}\\
    b^\mathrm{est} &= \frac{\sum_j\left(D_j-N_D^{-1}\sum_k D_k\right)\left(\overline Z_j-N_D^{-1}\sum_k\overline Z_k\right)}{\sum_j\left(D_j-N_D^{-1}\sum_k D_k\right)^2}. 
\label{eqn:b}
\end{align}
The estimators defined by Eqs.~\eq{eqn:a} and \eq{eqn:b} are Gaussian random variables with $E(a^\mathrm{est})=a$, $E(b^\mathrm{est})=b$, and
\begin{align*}
    \var(a^\mathrm{est})=\frac{N_D^{-1}\sum_j D_j^2}{\sum_j\left(D_j-N_D^{-1}\sum_k D_k\right)^2}\var(\overline Z), \qquad 
    \var(b^\mathrm{est}) =\frac{\var(\overline Z)}{\sum_j\left(D_j-N_D^{-1}\sum_k D_k\right)^2} , \\
    \cov(a^\mathrm{est},b^\mathrm{est})\equiv {\mathbb E} [(a^\mathrm{est}-a)(b^\mathrm{est}-b)]=
        -\frac{N_D^{-1}\sum_j D_j}{\sum_j\left(D_j-N_D^{-1}\sum_k D_k\right)^2}\var(\overline Z) .
\end{align*}
To estimate $\mu_B$, we use
\begin{align}
\tilde\mu_B^\mathrm{est}\equiv a^\mathrm{est}/b^\mathrm{est}.   
\label{eqn:muBhat}
\end{align}
When $\var(a^\mathrm{est})^{1/2}$ and $\var(b^\mathrm{est})^{1/2}$ are sufficiently small, we expand $\tilde\mu_B^\mathrm{est}$ in powers of $\delta a^\mathrm{est}\equiv a^\mathrm{est}-a$ and $\delta b^\mathrm{est}\equiv b^\mathrm{est} -b$. Truncating to the first order in $\delta a^\mathrm{est}$ and $\delta b^\mathrm{est}$, $\tilde\mu_B^\mathrm{est}$ becomes approximately Gaussian-distributed with ${\mathbb E}(\tilde\mu_B^\mathrm{est})\approx \mu_B$ and
\begin{align}
\var(\tilde\mu_B^\mathrm{est})\approx \frac{b^2\,\var(a^\mathrm{est})+a^2\,\var(b^\mathrm{est})-2ab\,\cov(a^\mathrm{est},b^\mathrm{est})}{b^4}. \label{eqn:varmuB}
\end{align}
For the experimental data presented in Fig.~\ref{fig:Fig4}a of the main text, $\var(\tilde\mu_B^\mathrm{est})$ is estimated by replacing $a\rightarrow a^\mathrm{est}$ and $b\rightarrow b^\mathrm{est}$ in Eq.~\eq{eqn:varmuB}, with $a^\mathrm{est}$ and $b^\mathrm{est}$ given by Eqs.~\eq{eqn:a} and \eq{eqn:b}, respectively.

Finally, to isolate the shift in the qubit frequency due to the first cumulant of the engineered source of noise, we apply the above procedure first in the absence of noise, and then in its presence. This yields the two sets of data points shown in Fig.~\ref{fig:Fig4}a of the main text. Fitting a straight line through each data set and substracting their $x$-intercept then gives our final estimate of $\mu_B$,
$\mu_B^\mathrm{est}=\mu_B^\mathrm{on}-\mu_B^\mathrm{off},$
where $\mu_B^\mathrm{on(off)}$ is the estimator defined by Eq.~\eq{eqn:muBhat} with ($\mu_B^\mathrm{on}$) or without ($\mu_B^\mathrm{off}$) engineered noise. The variance of $\mu_B^\mathrm{est}$ is then simply $\var(\mu_B^\mathrm{est})=\var(\mu_B^\mathrm{on})+\var(\mu_B^\mathrm{off})$. For the experimental data presented in the main text, we find $\mu_B^\mathrm{est}/2\pi=127.1$~kHz with a standard deviation $\var(\mu_B^\mathrm{est})^{1/2}=3.86$~kHz, corresponding to the 95\% confidence interval $\mu_B^\mathrm{est}/2\pi=127.1\,\pm\,7.56$~kHz.

\vspace*{8mm}

\subsection{PSD estimation procedure}
\label{sec:psd}

To estimate the PSD, we build on the frequency-comb approach introduced by Alvarez and Suter in Ref. \cite{Alvarez2011}. As detailed in the main text, treating the FFs as frequency combs generates a system of linear equations, solving which determines the PSD sampled at the harmonic frequencies. Rather than solving this sytem by matrix inversion as in Ref. \cite{Alvarez2011}, we employ a statistically-motivated maximum likelihood estimate (MLE), which takes experimental error into account. The likelihood function we use follows from the asymptotic Gaussian distribution of the measured decay constants, which we describe in Sec. \ref{sec::chi}. In Sec. \ref{sec::PSDMLE}, we determine the likelihood, the conditional probability of obtaining a set of decay data conditioned on the actual value of the PSD. The task of maximizing the likelihood can be cast as a linear problem, allowing clear comparison with Ref. \cite{Alvarez2011}.

\subsubsection{Distribution of the decay constant 
\label{sec::chi}}

The PSD enters the dynamics of the qubit through the decay constant in Eq. (\ref{eqnchi}), which can be obtained from measurements of the transverse Pauli operators, $\sigma_x$ and $\sigma_y$. Let $\sigma_i^\text{est}$ denote the estimated expected value of $\sigma_i$ for $i\in\{x,y\}$ after the qubit has evolved for a time $t$ under control sequence $p$. In the limit of a large number of measurements, $\sigma_i^\text{est}$ is approximately Gaussian distributed with mean $\mu_i=\mathbb{E}[\langle\sigma_i(t)\rangle]$ and variance $\text{var}[\sigma_i^\text{est}]$. In terms of the estimated expected values, the estimated decay constant is
\begin{align}\label{eq::ChiEst}
\chi_p^\text{est}(t)=-\frac{1}{2}\text{ln}(\sigma_x^{\text{est}\;2}+\sigma_y^{\text{est}\;2})
=-\frac{1}{2}\text{ln}\big[(\tilde{\sigma}_x^{\text{est}}+\mu_x)^2+(\tilde{\sigma}_y^{\text{est}}+\mu_y)^2\big],	
\end{align}
where $\tilde{\sigma}_i^{\text{est}}=\sigma_i^{\text{est}}-\mu_i$. Note that ${\sigma}_i^{\text{est}}$, $\mu_i$ and $\tilde{\sigma}_i^{\text{est}}$ on the right side of this expression depend implicitly on the time $t$.
When $\text{var}[\sigma_x^\text{est}],\,\text{var}[\sigma_y^\text{est}]\ll 1$, we can expand this expression about 
$\tilde{\sigma}_x^{\text{est}},\tilde{\sigma}_y^{\text{est}}\approx 0$, yielding
\begin{align*}
\chi_p^\text{est}(t)\approx-\frac{1}{2}\text{ln}(\mu_x^2+\mu_y^2)-\left(\frac{\mu_y}{\mu_x^2+\mu_y^2}\right)\tilde{\sigma}_y^{\text{est}}-\left(\frac{\mu_x}{\mu_x^2+\mu_y^2}\right)\tilde{\sigma}_x^{\text{est}}.
\end{align*}
Since it is a linear combination of Gaussian distributed random variables, the decay constant is also Gaussian distributed with mean and variance,
\begin{align}\label{eq::Echi}
{\mathbb E}[\chi_p^\text{est}(t)]=&-\frac{1}{2}\text{ln}(\mu_x^2+\mu_y^2)=\frac1{2\pi}\int_{-\infty}^\infty d\omega |F_p(\omega,t)|^2S(\omega)+
\Theta(t^4)\\\label{eq::Vchi}
\text{var}[\chi_p^\text{est}(t)]=&
\left(\frac{\mu_y}{\mu_x^2+\mu_y^2}\right)^{\!2}\text{var}[{\sigma}_y^{\text{est}}]
+\left(\frac{\mu_x}{\mu_x^2+\mu_y^2}\right)^{\!2}\text{var}[{\sigma}_x^{\text{est}}].
\end{align}

\subsubsection{Maximum likelihood estimate\label{sec::PSDMLE}}
Recall from the main text that after $M\gg 1$ repetitions of a control sequence $p$ with cycle time $T$, the FF in Eq.  (\ref{eq::Echi}) is an approximate frequency comb, enabling us to write the decay constant as 
\begin{align*}
{\mathbb E}[\chi_p^\text{est}(MT)]\approx\frac{M}{T}\sum_{k=-\infty}^\infty|F_p(k\omega_h,T)|^2S(k\omega_h).    
\end{align*}
Using the symmetry $S(\omega)=S(-\omega)$, and the decay of the PSD and FF at high frequencies, we can truncate the sum above to a finite number of harmonic frequencies, 
\begin{align*}
{\mathbb E}[\chi_p^\text{est}(MT)]\approx\frac{M}{T}\sum_{k=0}^{K-1}\Big(\frac{2-\delta_{k,0}}{2}\Big)|F_p(k\omega_h,T)|^2S(k\omega_h).    
\end{align*}
From the expected value above and the variance in Eq. (\ref{eq::Vchi}), the conditional probability of measuring $\chi_p^\text{est}(MT)$ given the actual values of the PSD, $\vec{S}=[S(0),\ldots,S(K\omega_h)]^T$, is 
\begin{align}
\label{eq::CondPchi}
P\big[\chi_p^\text{est}(MT)|\vec{S}\big]=
\frac{1}{\sqrt{2\pi\,\text{var}[\chi_p^\text{est}(MT)]}}\exp \left\{ -\frac{\Big [ \chi_p^\text{est}(MT) - \frac{M}{T}\sum_{k=0}^{K-1}\Big(\frac{2-\delta_{k,0}}{2}\Big)|F_p(k\omega_h,T)|^2S(k\omega_h) \Big]^2}{2\,\text{var}[\chi_p^\text{est}(MT)]}   \right\}.
\end{align}
The estimate of the PSD is based on the likelihood or conditional probability of measuring $\vec{\chi}=[\chi_1^\text{est}(MT),\ldots,\chi_P^\text{est}(MT)]^T$ for a set of control sequences $p=1,\ldots,P$ with $P\geq K$. Since the measurements of the decay constant are uncorrelated, the likelihood follows from Eq. (\ref{eq::CondPchi}),
\begin{align}
\label{eq::chilikelihood}
P\big(\vec{\chi} | \vec{S}\big) = \prod_{p=1}^{P} P\big[\chi_p^\text{est}(MT) | \vec{S}\big] =\left[ (2\pi)^{P} \det \mathbf{\Sigma} \right]^{-\frac{1}{2}} \text{exp} \left[ -\frac{1}{2} (\vec{\chi} - \mathbf{B} \vec{S})^T \mathbf{\Sigma}^{-1} (\vec{\chi} - \mathbf{B} \vec{S} )\right],
\end{align}
where the $P\times P$ covariance matrix has elements $\mathbf{\Sigma}_{p,q} = \text{var}[\chi_p^\text{est}(MT)]\,\delta_{p,q}$ and the $P\times K$ reconstruction matrix $\mathbf{B}$ depends on the  FFs evaluated at the harmonic frequencies
\begin{align*}
\mathbf{B}_{p,k}=\frac{M}{T}\Big(\frac{2-\delta_{k,0}}{2}\Big)|F_p(k\omega_h,T)|^2.
\end{align*}
Since the likelihood is Gaussian, the maximum likelihood estimate of the PSD is equivalent to the value of $\vec{S}$ that minimizes the argument of the exponential in Eq. (\ref{eq::chilikelihood}),
\begin{align}\label{eq::chiMLE}
\vec{S}^{\,\text{MLE}}=\argmin_{\vec{S}}\frac{1}{2} (\vec{\chi} - \mathbf{B} \vec{S})^T \mathbf{\Sigma}^{-1} (\vec{\chi} - \mathbf{B} \vec{S} ).
\end{align}
The least squares estimate of the PSD originally used in Ref. \cite{Alvarez2011}, given by $\vec{S}^\text{\,LS}=\mathbf{B}^{-1}\vec{\chi}$, is recovered when $\mathbf{\Sigma}=\mathbf{I}$. This implies that all measurements of the decay constant contribute equally to the estimate. In contrast, the dependence of $\vec{S}^{\,\text{MLE}}$ on the actual variances in $\mathbf{\Sigma}$ ensures that measurements with more uncertainty contribute less to the estimate.

In the experiment, the PSD is reconstructed at the harmonics $k=0,\ldots,7$ using the $P=11$ control sequences depicted in Fig. (\ref{fig:SuppFig3}), all with cycle time $T=960\,$ns. This differs from the implementation of Ref. \cite{Alvarez2011}, which uses CPMG control sequences of varying cycle times. For each control sequence, the transverse Pauli components are measured to obtain $\sigma_x^\text{est}$, $\sigma_y^\text{est}$ and $\chi_p^\text{est}(MT)$ from Eq. (\ref{eq::ChiEst}). The variances of $\chi_p^\text{est}(MT)$, which comprise $\mathbf{\Sigma}$, follow from Eq. (\ref{eq::Vchi}) with $\mu_x$ and $\mu_y$ replaced by the estimated values $\sigma_x^\text{est}$ and $\sigma_y^\text{est}$. In principle, the use of $P=11$ control sequences would enable us to reconstruct the PSD at $K=11$ harmonics. For the particular set of control sequences we used, however, the reconstruction matrix $\mathbf{B}$ becomes ill-conditioned for $K>8$, limiting the number of reconstructable harmonics.

\subsection{Bispectrum estimation procedure}   

 \label{sec:estimation:procedure}
\newcommand{\matr}[1]{\mathbf{#1}} 
\newcommand{\norm}[1]{\left\lVert#1\right\rVert_2} 
\newcommand{\argmax}{\mathop{\mathrm{argmax}}} 
\newcommand{\mean}[1]{{\mu_{#1}}} 
\newcommand{\std}[1]{\mathrm{\sigma_{#1}}} 
\newcommand{\tr}[1]{{#1}^{\top}} 

While this work is based on the non-Gaussian QNS protocol originally proposed in Ref. \cite{Norris2016}, the estimation procedure we implemented contains several innovations aimed at generalizing the noise model and improving robustness in the presence of experimental error and numerical instability. First, the zero-mean, non-Gaussian noise model of Ref. ~\cite{Norris2016} is insufficient to describe
the square noise engineered in our quarton-qubit sensor, which is inherently nonzero mean. This complicates the estimation procedure, since both the bispectrum and the mean enter the qubit dynamics through the phase in Eq. (\ref{eqnphi}). Estimating the bispectrum requires that we disambiguate the phase contribution of the bispectrum from that of the mean, which we accomplish by 
 first estimating the noise mean and then isolating the dynamical contribution of the bispectrum in the non-Gaussian phase. A second key difference is the ``single-shot" nature of the current estimation procedure. In Ref. \cite{Norris2016}, control sequences with non-zero filter order were first used to estimate the bispectrum at ``non-zero harmonics", i.e., $(k_1\omega_h,k_2\omega_h)$ for which $k_1,k_2\neq 0$. This estimate was combined with subsequent phase measurements to estimate the bispectrum at ``zero harmonics", i.e. $(k_1\omega_h,k_2\omega_h)$ for which $k_1=0$ and/or $k_2=0$. In the RMLE estimate of the spectrum presented here, both the zeros and the non-zero harmonics are estimated simultaneously, eliminating any compounding of error that can occur in the two-step procedure. The present work also departs from Ref. \cite{Norris2016} significantly in its use of a statistically motivated maximum likelihood estimation procedure. As discussed in the main text, the least-squares estimate of Ref. \cite{Norris2016} is susceptible to numerical instability and, additionally, does not take measurement error into account.
 
In the remainder of this section, we fully detail our bispectrum estimation procedure. 
We begin by describing the probability distribution of the estimated ``non-Gaussian'' phase angle ($\varphi_p$), which enables us to derive the likelihood function for the probability of obtaining a particular set of phase data conditioned on the actual value of the bispectrum (Subsection~\ref{subsec:Likelihood}). From the likelihood, 
the task of reconstructing the bispectrum can be mapped into an RMLE problem, as shown Subsection~\ref{subsec:RMLE}. The RMLE approach increases numerical stability, accounts for experimental error and allows us to deploy prior knowledge of the bispectrum in the estimation procedure. Since regularization can introduce error into the estimate if it is too strong, in Subsection~\ref{subsec:L-curve} we determine an appropriate regularization strength for our problem using the L-curve criterion. 

\subsubsection{Distribution of the non-Gaussian phase}
\label{subsec:Distribution_tilde_phi}
The non-Gaussian phase of the qubit is determined from the  estimated expected values of the transverse Pauli operators, $\sigma_x^\text{est}$ and $\sigma_y^\text{est}$, when the qubit has evolved under control sequence $p$ for a time $t$. Recall from Sec. \ref{sec:psd} that in the limit of a large number of measurements, $\sigma_i^\text{est}$ is approximately Gaussian distributed with mean $\mu_i=\mathbb{E}[\langle\sigma_i(t)\rangle]$ and variance $\text{var}[\sigma_i^\text{est}]$. From the estimated expected values, the ordinary phase is determined by
\begin{align}\label{eq::PhiEst}
\phi_p^\text{est}(t) =- \tan^{-1} \left( \frac{\sigma_x^{\text{est}}}{\sigma_y^{\text{est}}} \right) = - \tan^{-1} \left( \frac{\tilde{\sigma}_x^{\text{est}}+\mean{x}}{\tilde{\sigma}_y^\text{est}+\mean{y}} \right),
\end{align}
where ${\sigma}_i^\text{est}$, $\mean{i}$ and $\tilde{\sigma}_i^\text{est} = \sigma_i^\text{est} - \mean{i}$ depend implicitly on the time $t$. When $\text{var}[\sigma_x^\text{est}]$, $\text{var}[\sigma_y^\text{est}] \ll 1$, we can expand $\phi^\text{est}_p$ about $\tilde{\sigma}_x^\text{est}$, $\tilde{\sigma}_y^\text{est}$ $\approx 0$, yielding
\begin{align*}
\phi_p^\text{est}(t) \approx -\tan^{-1} \left( \frac{\mean{x}}{\mean{y}} \right) - \left( \frac{\mean{y}}{\mu_x^2 + \mu_y^2} \right) \tilde{\sigma}_x^\text{est} + \left( \frac{\mean{x}}{\mu_x^2 + \mu_y^2} \right) \tilde{\sigma}_y^\text{est}.
\end{align*} 
As a linear combination of Gaussian distributed random variables, the phase is also Gaussian distributed with mean and variance  
\begin{align}\label{eq::Ephi}
{\mathbb E}[\phi_p^\text{est}(t)] &= -\tan^{-1} \left( \frac{\mean{x}}{\mean{y}}\right) = F_p(0,t)\mu_B -\frac1{3!(2\pi)^2}\int_{\mathbb R^2} d\vec \omega \,G_p(\vec\omega,t)\,S_2(\vec\omega)+\Theta[t^5], \\
\text{var}[\phi_p^\text{est}(t)] &= \left(\frac{\mu_y}
{\mu_x^2 + \mu_y^2}\right)^{\!2}\text{var}[\sigma_x^\text{est}] + \left(\frac{\mu_x}{\mu_x^2 + \mu_y^2}\right)^{\!2}\text{var}[\sigma_y^\text{est}].\notag
\end{align}
The second equality on the right-hand side of $\mathbb{E}[\phi_p^\text{est}]$ follows from Eq. (\ref{eqnphi}). Subtracting out the contribution of the noise mean from the phase produces the non-Gaussian phase,
\begin{align}
\varphi_p^\text{est}(t)  =  \phi_p^\text{est}(t) - F_p(0,t) \mu_B^\text{est},
\label{eq::PhiMod}
\end{align}
where $\mu_B^\text{est}$ is the estimated noise mean described in Sec. \ref{sec:meanEstimate}. 
Using the asymptotic Gaussian distribution of $\mu_B^\text{est}$ with mean $\mu_B$ and variance $\text{var}[\mu_B^\text{est}]$,  the non-Gaussian phase is similarly Gaussian with mean and variance
\begin{align}
{\mathbb E}[\varphi_p^\text{est}(t)] &= - \frac1{3!(2\pi)^2}\int_{\mathbb R^2} d\vec \omega \,G_p(\vec\omega,t)\,S_2(\vec\omega)+\Theta[t^5],
\label{eq::EPhiT}\\
\text{var}[\varphi_p^\text{est}(t)] &= \left(\frac{\mu_y}{\mu_x^2 + \mu_y^2}\right)^{\!2}\text{var}[\sigma_x^\text{est}] + \left(\frac{\mu_x}{\mu_x^2 + \mu_y^2}\right)^{\!2}\text{var}[\sigma_y^\text{est}]+F_p(0,t)^2\,\text{var}[\mu_B^\text{est}].
\label{eq::VarPhiT}
\end{align}
Note that $\mathbb{E}[\varphi_p^\text{est}(t)]$ depends to leading order on the bispectrum, unlike $\mathbb{E}[\phi_p^\text{est}(t)]$ above.

\subsubsection{Restriction to the principal domain}
\label{subsec:PD}
For any real, classical process, the bispectrum has three general symmetries: (1) $S_2(\omega_1,\omega_2)=S_2(\omega_2,\omega_1)$ (permutation symmetry); (2) $S_2(\omega_1,\omega_2)=S_2(-\omega_1,-\omega_2)$ (invariance under complex conjugation); (3) $S_2(\omega_1,\omega_2)=S_2(-\omega_1-\omega_2,\omega_2)$ (stationarity). These symmetries define the 12 regions of the frequency plane depicted in Fig. \ref{fig:Fig3}c. If $(\omega_1,\omega_2)\in\text{int}(\mathcal{D}_2)$ is contained in the interior of the principal domain, the symmetries imply
\begin{align*}
S_2(\omega_1,\omega_2)=S_2(\omega_2,\omega_1)=
S_2(-\omega_2,\omega_1+\omega_2)=S_2(-\omega_1,\omega_1+\omega_2)=
S_2(-\omega_1-\omega_2,\omega_1)=S_2(-\omega_1-\omega_2,\omega_2)=\\
S_2(-\omega_1,-\omega_2)=S_2(-\omega_2,-\omega_1)=S_2(\omega_2,-\omega_1-\omega_2)=S_2(\omega_1,-\omega_1-\omega_2)
=S_2(\omega_1+\omega_2,-\omega_1)=S_2(\omega_1+\omega_2,-\omega_2).
\end{align*}
In other words, the bispectrum takes a value equivalent to $S_2(\omega_1,\omega_2)$ in each of the 12 regions. This is summarized by the multiplicity, $m(\omega_1,\omega_2)=12$. For $(\omega,\omega)$, which lies on the boundary of $\mathcal{D}_2$, 
\begin{align*}
S_2(\omega,\omega)=S_2(-\omega,-\omega)=S_2(-2\omega,\omega)=
S_2(\omega,-2\omega)=S_2(2\omega,-\omega)=S_2(-\omega,2\omega),
\end{align*}
implying $m(\omega,\omega)=6$. For $(\omega,0)$, which also lies on the boundary of $\mathcal{D}_2$,
\begin{align*}
S_2(\omega,0)=S_2(0,\omega)=S_2(-\omega,0)=S_2(0,-\omega)=S_2(\omega,-\omega)=S_2(\omega,-\omega),
\end{align*}
similarly implying $m(\omega,0)=6$. Note that $(\omega_1,\omega_2)=(0,0)$ is invariant under all of the symmetries, implying that $m(0,0)=1$.

The symmetries and multiplicities simplify the expected value of the phase substantially.
Recall that after $M\gg 1$ repetitions of control sequence $p$ with cycle time $T$, the frequency comb approximation enables us to write the expected phase as a discrete sum depending on the bispectrum and the FF evaluated at the harmonic frequencies,
\begin{align*}
{\mathbb E}[\varphi_p^\text{est}(MT)]=-\frac1{3!(2\pi)^2}\int_{\mathbb R^2} d\vec \omega \,G_p(\vec\omega,MT)S_2(\vec\omega) \approx -\frac{M}{3!T^2} \sum_{\vec{k} \in \mathbb{Z}^2} G_p(\omega_h\vec{k},T)  S_2(\omega_h\vec{k}).
\end{align*}
In terms of the multiplicities, we can rewrite the sum as 
\begin{align*}
{\mathbb E}[\varphi_p^\text{est}(MT)]\approx&-\frac{12M}{3!T^2} \sum_{\omega_h\vec{k} \in \text{int}(\mathcal{D}_2)} G_p(\omega_h\vec{k},T)  S_2(\omega_h\vec{k})-\frac{6M}{3!T^2} \sum_{k\in\mathbb{Z}} G_p(\omega_hk,\omega_hk,T)  S_2(\omega_hk,\omega_hk)\\
&-\frac{6M}{3!T^2} \sum_{k\in\mathbb{Z}} G_p(\omega_hk,0,T)  S_2(\omega_hk,0)-\frac{M}{3!T^2} G_p(0,0,T)  S_2(0,0)\\
=&-\frac{M}{3!T^2} \sum_{\omega_h\vec{k} \in \mathcal{D}_2} m(\omega_h\vec{k})  G_p(\omega_h\vec{k},T) S_2(\omega_h\vec{k}).
\end{align*}
Using $S_2(\omega_h\vec{k})=S_2(-\omega_h\vec{k})$ and $G_p(\omega_h\vec{k},T)^*=G_p(-\omega_h\vec{k},T)$, and truncating the sum to a finite subset $\mathcal{K}_2$, we obtain
\begin{align}
{\mathbb E}[\varphi_p^\text{est}(MT)]\approx-\frac{M}{3!T^2} \sum_{\vec{k} \in \mathcal{K}_2} m(\omega_h\vec{k})  \text{Re}[G_p(\omega_h\vec{k},T)] S_2(\omega_h\vec{k}).
\label{eq::C3Comb}
\end{align}

\subsubsection{Likelihood function $P(\vec{\varphi}| \vec{S}_2)$}
\label{subsec:Likelihood}
Given the actual bispectrum in $\mathcal{K}$, $\vec{S}_2=[S_2(\omega_h\vec{k}_1),\ldots,S_2(\omega_h\vec{k}_N)]^T$, the conditional probability of measuring $\varphi_p^\text{est}(MT)$  follows from Eqs. (\ref{eq::EPhiT})-(\ref{eq::C3Comb}),
\begin{align}
\label{eq::Pphi}
P\big[\varphi_p^\text{est}(MT)|\vec{S}_2\big] = \frac{1}{\sqrt{2\pi\,\text{var}[\varphi_p^\text{est}(MT)]}}\exp \left\{ -\frac{\Big [ \varphi_p^\text{est}(MT) + \frac{M}{3! T^2} \sum_{\vec{k}\in\mathcal{K}} m(\omega_h\vec{k}) \mathrm{Re} [G_p(\omega_h\vec{k},T)] S_2(\omega_h\vec{k}) \Big]^2}{2\,\text{var}[\varphi_p^\text{est}(MT)]} \right\}.
\end{align}
Reconstructing the bispectrum requires measurements the non-Gaussian phase  for a set of control sequences $p = 1,\ldots,P$ with $P\geq N$, which we gather  into the column vector $\vec{\varphi}=[\varphi_1^\text{est}(MT),\ldots,\varphi_{P}^\text{est}(MT)]^T$.  Because the non-Gaussian phase measurements are uncorrelated, the likelihood or probability of obtaining $\vec{\varphi}$ given $\vec{S}_2$ is a product of the conditional probabilities for the complete set of control sequences,
\begin{align}\label{eq::likelihood}
P\big(\vec{\varphi} | \vec{S}_2\big) = \prod_{p=1}^{P} P\big[\varphi_p^\text{est}(MT) | \vec{S}_2\big] =\left[ (2\pi)^{P} \det \matr{\Sigma} \right]^{-\frac{1}{2}} \exp \left[ -\frac{1}{2} (\vec{\varphi} - \matr{A} \vec{S}_2)^T \matr{\Sigma}^{-1} (\vec{\varphi} - \matr{A} \vec{S}_2 )\right].
\end{align}
Here, the $P \times P$ covariance matrix $\matr{\Sigma}$ is diagonal with elements $\matr{\Sigma}_{p,q} = \text{var}[\varphi_p^\text{est}(MT)]\,\delta_{p,q}$, and the $P\times N$ reconstruction matrix 
$\matr{A}$ depends on the filter functions evaluated at the harmonic frequencies,
\begin{align}\label{eq::Apn}
(\matr{A})_{p,n} = -\frac{M}{3!T^2}\, m(\omega_h\vec{k}_n)\, \mathrm{Re} [ G_p(\omega_h\vec{k}_n,T)].
\end{align}

In the experiment, the likelihood in Eq. (\ref{eq::likelihood}) is determined by measuring the non-Gaussian phase for each of the $P=11$ control sequences in Fig. (\ref{fig:SuppFig3}). Since we also rely on these sequences to estimate the PSD, both the bispectrum and the PSD can be estimated with the the same set of transverse Pauli measurements. For each of the control sequences, $\varphi_p^\text{est}(MT)$ was determined from Eq. (\ref{eq::PhiEst}) using measurements of $\sigma_x^\text{est}$, $\sigma_y^\text{est}$ and $\mu_B^\text{est}$. The variances of the $\varphi_p^\text{est}(MT)$, which constitute the covariance matrix $\matr{\Sigma}$, are given by Eq. (\ref{eq::VarPhiT}) with $\mu_x$ and $\mu_y$ replaced by the estimated values $\sigma_x^\text{est}$, $\sigma_y^\text{est}$.  The reconstruction matrix $\matr{A}$ is determined from Eq. (\ref{eq::Apn}), with each FF evaluated on the set of harmonics $\mathcal{K}_1$ depicted in Fig. \ref{fig:Fig3}(b).

\subsubsection{Regularized maximum likelihood estimation}
\label{subsec:RMLE}

For the Gaussian likelihood derived in the previous section, the maximum likelihood estimate (MLE) of the bispectrum is equivalent to the value of $\vec{S}_2$ that minimizes the exponent in Eq. (\ref{eq::likelihood}),  
\begin{align}\label{eq::MLE}
\vec{S}_2^{\,\text{MLE}}=  \argmin_{\vec{S}_2}\,\frac{1}{2}\big(\vec{\varphi} - \matr{A} \vec{S}_2\big)^T\, \matr{\Sigma}^{-1}\, \big(\vec{\varphi} - \matr{A} \vec{S}_2\big).
\end{align}
 In the special case where $\matr{\Sigma}\propto \matr{I}$, we recover the least-squares estimate used in Ref. \cite{Norris2016} with solution $\vec{S}_2^\text{\,LS}=\matr{A}^{-1}\vec{\varphi}$. Even with a nonuniform covariance matrix, Eq. (\ref{eq::MLE}) is a simple convex optimization problem admitting an analytic solution for $\vec{S}_2^{\,\text{MLE}}$.
When $\matr{A}$ is ill-conditioned, however, the MLE suffers from numerical instability, which can introduce significant error into the estimate of the bispectrum, despite the existence of an analytic solution. The problem can be made more stable by introducing a regularization term or ``regularizer"  $Q(\vec{S}_2)$,
producing the regularized maximum likelihood estimate (RMLE) of the bispectrum,
\begin{align}\label{eq::RMLE}
\vec{S}_2^{\,\text{RMLE}} = \argmin_{\vec{S}_2} \left[\frac{1}{2} (\matr{A} \vec{S}_2 - \vec{\varphi})^T \matr{\Sigma}^{-1} (\matr{A} \vec{S}_2 - \vec{\varphi}) + Q(\vec{S}_2) \right].
\end{align}
The regularizer imposes additional structure on the solution, making it more robust to numerical instability arising from $\matr{A}$. It also prevents overfitting, in which the estimated bispectrum is unduly influenced by errors in $\vec{\varphi}$ and is, thus, a poor predictor of the qubit dynamics under more general control settings.

\begin{figure}[h!]
\centering
\includegraphics{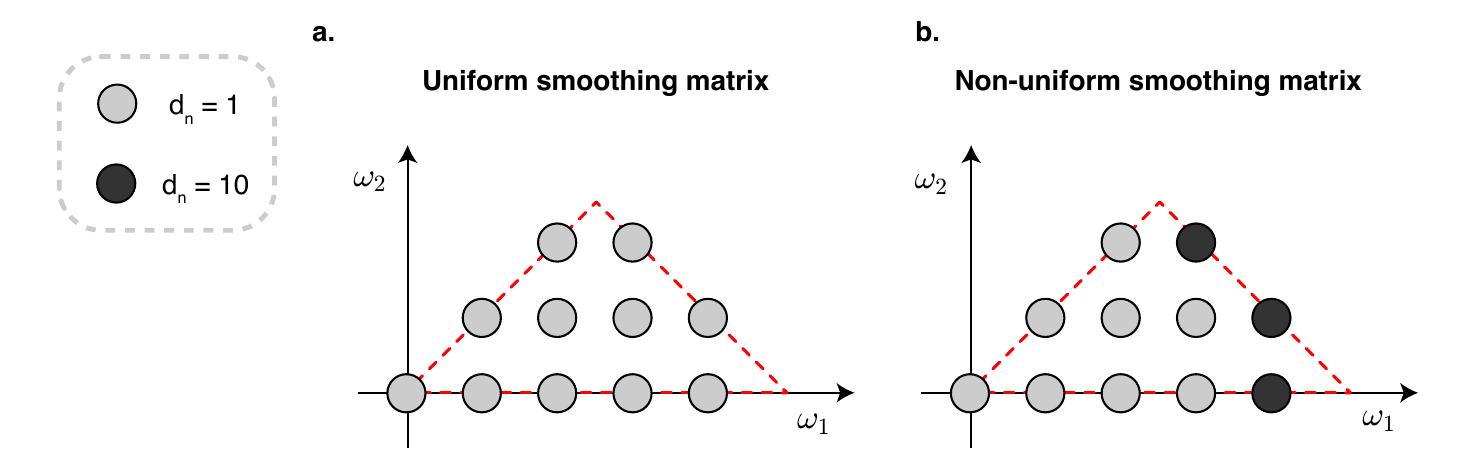}
\caption{
\textbf{Smoothing matrix $\matr{D}$ given a prior information.} 
\textbf{a,} A smoothing matrix that assumes the least prior information about the bispectrum.  
\textbf{b,} A smoothing matrix that assumes the bispectrum decays to zero at the border of an octant.}
\label{fig:SuppFig4_1}
\end{figure}

There are numerous methods of regularization for ill-conditioned and/or ill-posed problems. An approach particularly ammenable to maximum likelihood estimation is Tikhonov regularization, which employs an L2 regularizer $ Q(\vec{S}_2) =  |\!|\lambda\vec{S}_2|\!|_2^2$ with strength controlled by the regularization parameter $\lambda\geq0$ [S5]
. In Eq. (\ref{eq::RMLE}), this regularizer has the effect of penalizing $\vec{S}_2$ with larger $L_2$-norms. To estimate the bispectrum, we consider a variation of Tikhonov regularization in which
\begin{align}
\label{eq::regularizer}
Q(\vec{S}_2) = \big|\!\big| \lambda \matr{D} \vec{S}_2\big|\!\big|_2^2,
\end{align}
where $\matr{D} = \mathrm{diag}~(d_1, \cdots, d_{N})$ is the diagonal ``smoothing matrix''. Note that the Tikhonov regularizer is recovered when $d_1=\ldots=d_{N}=1$ (Fig.~\ref{fig:SuppFig4_1}a) and  the standard maximum likelihood estimate is recovered when $\lambda =0$. Using non-uniform values for the diagonals enables us to incorporate prior information about the bispectrum. For example, if the magnitude of the bispectrum is known to decay at the high-frequency border of $\mathcal{K}$, we can make the corresponding harmonics in $\matr{D}$ large compared to those of the interior (Fig.~\ref{fig:SuppFig4_1}b). Such a smoothing matrix  favors a solution with small magnitude at the border. The connection between the smoothing matrix and prior knowledge of the bispectrum is more explicit in a Bayesian formulation of the estimation problem in which the RMLE estimate in Eq. (\ref{eq::RMLE}) with the regularizer in Eq. (\ref{eq::regularizer}) is equivalent to a posterior mean estimate in which the prior distribution of the bispectrum is Gaussian and zero-mean with covariance matrix $(2\lambda^2\matr{D}^2)^{-1}$, provided $\matr{D}$ is full-rank. For any smoothing matrix, the regularized maximum likelihood estimate has the simple analytic solution,
\begin{align}
\vec{S}_2^{\,\text{RMLE}} = \left(\matr{A}^T \matr{\Sigma}^{-1} \matr{A} + 2 \lambda^2\matr{D}^2 \right)^{-1} (\matr{A}^T \matr{\Sigma}^{-1} \vec{\varphi}).
\label{eq:s_RMLE}
\end{align}

\subsubsection{The L-curve criterion}
\label{subsec:L-curve}

Although it guards against numerical instability and overfitting to errors in the measured data, the regularizer can introduce its own error into the estimate if $\lambda$ too large. A fundamental challenge in regularization is selecting a value of $\lambda$ that balances these sources of error. While this problem is still an active area of research, one of the most widely used strategies for selecting $\lambda$ is the L-curve criterion [S6].
A graphical technique, the L-curve criterion enables one to visualize the magnitude of the regularization error in proportion to other errors in the estimate and choose $\lambda$ accordingly.

\begin{figure}[t!]
\centering
\includegraphics{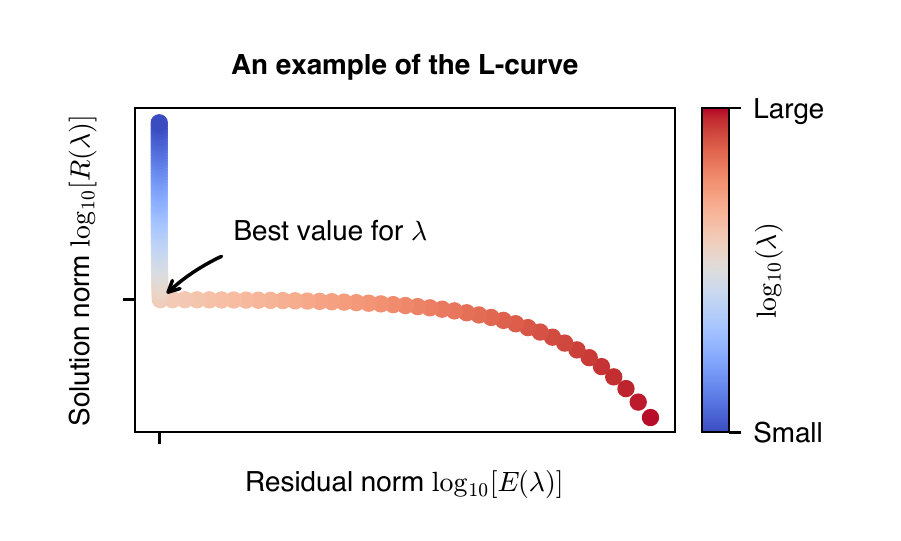}
\vspace*{-3mm}
\caption{\textbf{An example of the L-curve plot.}}
\label{fig:SuppFig4_2}
\end{figure}

For a given $\lambda$, the sources of error that contribute to the regularized maximum likelihood estimate in Eq. (\ref{eq:s_RMLE}) are described by the residual norm
\begin{align}
E(\lambda) \equiv \left[ \frac{1}{2}(\matr{A} \vec{S}_2^{\,\text{RMLE}} - \vec{\varphi})^T\, \matr{\Sigma}^{-1} (\matr{A} \vec{S}_2^{\,\text{RMLE}} - \vec{\varphi}) \right]^{1/2}.
\label{eq:residual_norm}
\end{align}
and the solution norm
\begin{align}
R(\lambda) \equiv \norm{\matr{D} \vec{S}_2^{\,\text{RMLE}}} = \norm{\text{diag}(d_1, \cdots, d_{|\mathcal{R}|}) \vec{S}_2^{\,\text{RMLE}}}.
\end{align}
The regularization parameter $\lambda$ enters both the residual norm and solution norm implicitly through $\vec{S}_2^{\,\text{RMLE}}$. The residual norm increases as $\lambda$ grows, while the solution norm decreases. When $E(\lambda)$ is too large relative to $R(\lambda)$, the estimate does not account for the measured data due to error introduced by the regularization. Conversely, when $E(\lambda)$ is too small relative to $R(\lambda)$, the estimate is fits the measured data too closely, making it susceptible to overfitting and numerical error. The influence of $\lambda$ on the error conributions is captured by the L-curve, a parametric plot of $\text{log}R(\lambda)$ vs. $\text{log}E(\lambda)$ as a function of $\lambda$. A typical L-curve, with its characteristic ``L" shape, is shown in Fig.~\ref{fig:SuppFig4_2}. Note that as $\lambda$ increases from left to right, $\text{log}R(\lambda)$ sharply decreases and then plateaus, while $\text{log}E(\lambda)$ is initially stable followed by a rapid increase. The corner of the L-curve, marks a point at which the solution norm and residual norm are small simultaneously. The corner, thus, signifies the optimal value of $\lambda$ according to the L-curve criterion.

\begin{figure}[h!]
\centering 
\includegraphics{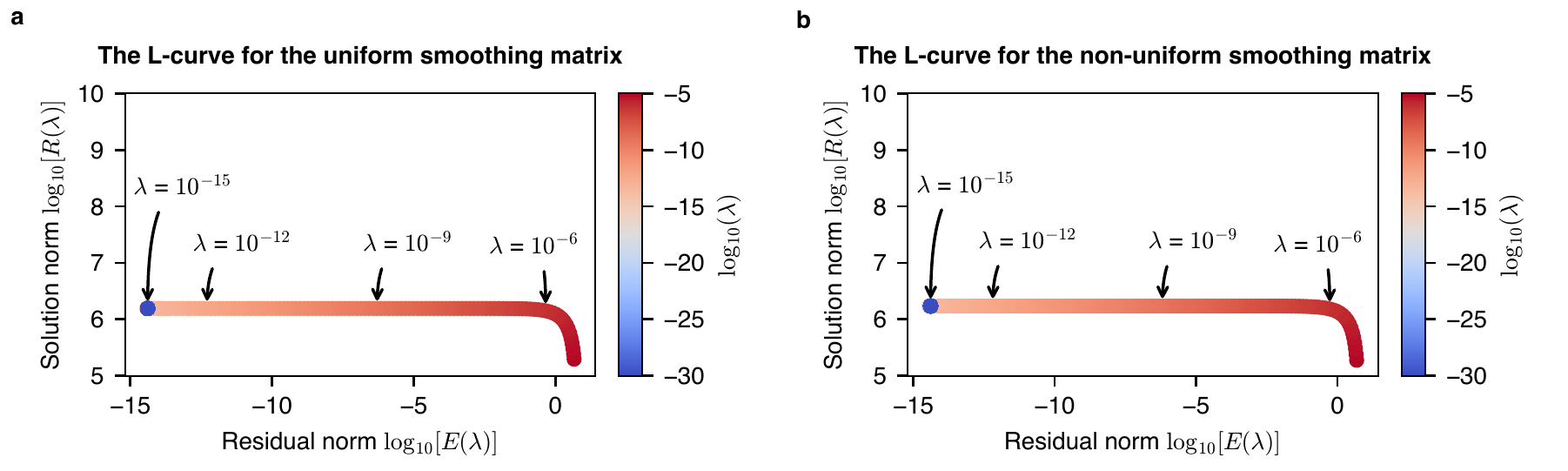}
\caption{
\textbf{The L-curve plots for experimental data.} 
\textbf{a,} The L-curve plot for the uniform smoothing matrix. 
\textbf{b,} The L-curve plot for the non-uniform smoothing matrix.}
\label{fig:SuppFig4_3}
\end{figure}

Figure ~\ref{fig:SuppFig4_3} shows L-curves generated by our experimental data for the two different smoothing matrices illustrated in Fig. ~\ref{fig:SuppFig4_1}. For both the uniform and non-uniform smoothing matrices, the L-curves lack corners. Unlike the typical L-curve in Fig.~\ref{fig:SuppFig4_2}, $\text{log}R(\lambda)$ does not exhibit a sharp increase as $\lambda\rightarrow 0$. This indicates that, for the control sequences we have selected, the reconstruction matrix $\matr{A}$ is sufficiently well conditioned to make regularization the dominant source of error [S7].
Consequently, it is  not optimal to utilize regularization in this setting and the reconstruction presented in the main text uses $\lambda=0$.   

Note that this finding is contingent on both $\matr{A}$ and the particular regularizer we employ. Estimating the bispectrum at a greater number of harmonics demands a larger $\matr{A}$, which is more likely to be near singular and/or poorly conditioned. This scenario will likely require some form of regularization. Additionally, the error introduced by regularization is reduced when prior knowledge of the bispectrum (if available) is used to select the regularizer. For example, suppose a noise model or previous experiment indicates that $\vec{S}_2$ takes a value in the vicinity of $\vec{S}_\mu$. This information is captured by the regularizer
\begin{align}
Q(\vec{S}_2) = \frac{1}{2}(\vec{S}_2-\vec{S}_\mu)^T(2\lambda^2\matr{D}^2)(\vec{S}_2-\vec{S}_\mu).
\end{align}
This corresponds to a Gaussian prior distribution of $\vec{S}_2$ with mean $\vec{S}_\mu$ and covariance matrix $(2\lambda^2\matr{D}^2)^{-1}$. In contrast, naively employing Tikhonov regularization amounts to assuming a zero-mean prior distribution of $\vec{S}_2$.

\newpage
\vspace*{1cm}
\subsection*{Additional references}

\begin{itemize}

\item[S1.] V. E. Manucharyan, J. Koch, L. I. Glazman, and M. H. Devoret, ``Fluxonium: Single Cooper-pair circuit free of charge offsets,'' Science \textbf{326}, 113--116 (2009).

\item[S2.] E. Magesan, J. M. Gambetta, and J. Emerson, ``Scalable and robust randomized benchmarking of quantum 
processes,'' Phys. Rev. Lett. {\bf 106}, 180504 (2011).

\item[S3.] C. Macklin, K. O'Brien, D. Hover, M. E. Schwartz, V. Bolkhovsky, X. Zhang, W. D. Oliver, and I. Siddiqi, 
``A near-quantum-limited Josephson traveling-wave parametric amplifier,'' Science {\bf 350}, 307--310 (2015).

\item[S4.] G. Casella and R. L. Berger, Statistical Inference (Duxbury Pacific Grove, CA, 2002).

\item[S5.] A. Tikhonov, ``Solution of incorrectly formulated problems and the regularization method,'' Dokl. Akad. Nauk {\bf 151}, 1035--1038 (1963).

\item[S6.] P. C. Hansen, ``The L-curve and its use in the numerical treatment of inverse problems,'' in
{\em Computational Inverse Problems in Electrocardiology} (WIT Press, 2000) pp. 119--142.

\item[S7.] T. Regi\'{n}ska, ``A regularization parameter in discrete ill- posed problems,'' SIAM J. Sci. Comput. {\bf 17}, 
740--749 (1996).
\end{itemize}

\end{document}